\documentclass{article}

\usepackage[final,nonatbib]{neurips_2023}

\usepackage[utf8]{inputenc} 
\usepackage[T1]{fontenc}    
\usepackage{hyperref}       
\usepackage{url}            
\usepackage{booktabs}       
\usepackage{amsfonts}       
\usepackage{nicefrac}       
\usepackage{microtype}      
\usepackage[table]{xcolor}         
\usepackage{setspace}
\usepackage{listings}
\usepackage{color}
\usepackage{soul}
\usepackage{tabularx}
\usepackage{enumitem}
\usepackage{algorithm, graphicx}
\usepackage{mdframed}
\usepackage{framed}
\usepackage{soul}
\usepackage{fontawesome}
\usepackage{tcolorbox}
\usepackage{verbatim}
\usepackage{mfirstuc}
\usepackage[flushleft]{threeparttable}
\usepackage{algpseudocode}

\usepackage[backend=biber, style=numeric, sorting=ynt]{biblatex}
\definecolor{mygreen}{rgb}{0,0.6,0}
\definecolor{mygray}{rgb}{0.5,0.5,0.5}
\definecolor{mymauve}{rgb}{0.58,0,0.82}
\lstdefinelanguage{ini}
{
	basicstyle=\ttfamily\footnotesize,
	columns=fullflexible,
	morecomment=[s][\color{Orchid}\bfseries]{[}{]},
	morecomment=[l]{\#},
	morecomment=[l]{;},
	commentstyle=\color{mygreen},
	morekeywords={DataClass,LargeClass,Blob},
	otherkeywords={=,:},
	keywordstyle={\color{blue!20}\bfseries}
}

\title{Empirical Investigation of the Relationship Between Design Smells and Role Stereotypes}

\author{ \href{https://orcid.org/0000-0002-0133-8164}{\includegraphics[scale=0.06]{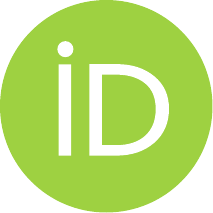}\hspace{1mm}Daniel Ogenrwot} \\
	Howard R. Hughes College of Engineering\\
        Department of Computer Science\\
	University of Nevada-Las Vegas\\
	Las Vegas, U.S.A \\
	\texttt{ogenrwot@unlv.nevada.edu} \\
	\And
	\href{https://orcid.org/0000-0002-0108-3798}{\includegraphics[scale=0.06]{orcid.pdf}\hspace{1mm}Joyce Nakatumba-Nabende}\thanks{Corresponding author} \\
	Department of Computer Science\\
	Makerere University\\
	Kampala, Uganda\\
	\texttt{jnabende@mak.ac.ug} \\
        \And
        \href{https://orcid.org/0000-0003-3206-7085}{\includegraphics[scale=0.06]{orcid.pdf}\hspace{1mm}John Businge}\\
	Howard R. Hughes College of Engineering\\
        Department of Computer Science\\
	University of Nevada-Las Vegas\\
	Las Vegas, U.S.A \\
	\texttt{john.businge@unlv.edu} \\
         \And
        \href{https://orcid.org/0000-0001-7517-6666}{\includegraphics[scale=0.06]{orcid.pdf}\hspace{1mm}Michel R.V. Chaudron}\\
	Department of Mathematics and Computer Science\\
	Eindhoven University of Technology\\
	Netherlands \\
	\texttt{m.r.v.chaudron@tue.nl} \\
}

\begin{document}
\maketitle
\vspace{-1cm}
\spacing{1.1}
\begin{abstract}
During the development of software systems, poor design and implementation choices can have a detrimental impact on the maintainability of the software. Design smells, recurring patterns of poorly designed fragments in software, are indicative of these issues. Role-stereotypes signify the generic responsibilities that classes assume in system design. Although the concepts of role-stereotypes and design smells are inherently different, both significantly contribute to the design and maintenance of software systems. Studying this relation is essential for (i) software maintainability, (ii) code quality improvement, (iii) efficient code review, (iv) guided refactoring, (v) early defect detection and (vi) role-specific metrics design.  This paper employs an exploratory approach, combining statistical analysis and unsupervised learning methods, to comprehend the relationship between design smells and role-stereotypes and how this connection varies across different desktop and mobile applications. The study utilizes a dataset comprising of 11,350 classes across 15 desktop and 15 mobile applications mined from GitHub. Overall, the findings reveal several design smells that co-occur more frequently across the entire role-stereotype categories. Specifically, three (3) out of six (6) role-stereotypes considered in this study are more susceptible to design smells. Furthermore, we examined the variation of design smells between desktop and mobile applications, driven by notable differences in their architectural paradigms. Mobile applications commonly rely on more loosely coupled UI frameworks and prioritize maintainability, attributed to frequent updates on app stores. Our findings revealed a higher prevalence of design smells in desktop applications compared to their mobile counterparts, particularly within the Service Provider and Information Holder role-stereotypes.
The result showed that design smells are more prevalent in desktop applications than in mobile applications, especially in the Service Provider and Information Holder role-stereotypes. Through unsupervised learning methods, it is observed that certain pairs or groups of role-stereotypes are prone to similar types of design smells compared to others. We believe that these relationships may be associated with the characteristic and collaborative properties between role-stereotypes. Therefore, this study offers crucial insights into previously undisclosed behavior regarding the relationship between design smells and role-stereotypes. The results of this paper can guide software teams in implementing various design smell prevention and correction mechanisms, as well as ensuring the conceptual integrity of classes during their design and maintenance. 
\end{abstract}

\onehalfspacing
\section{INTRODUCTION}\label{sec:intro}


Software design is an essential component of software engineering. A well-designed software system leads to reliable and maintainable software \cite{imran2019design, imran2022qualitative}. New software applications nowadays are quite complex and constantly adapting to the ever-changing user requirements, which poses a challenge of maintainability. The ever increasing maintenance costs \cite{jha2019deep} are often a consequence of bad design and development practices commonly observed in software systems. While developing software systems, poor design and implementation choices in source code can negatively affect the maintainability of software systems \cite{palomba2018diffuseness}. The problematic structures identified in source code are known as design smells \cite{kaur2015influence,aversano2020empirical,imran2022qualitative}. Design smells represent structural anomalies that deviate from established design principles, introducing technical debt and impeding the overall effectiveness of the software. For instance, a ``God class" assumes too many responsibilities and usually has very many methods, violating the single-responsibility principle. This design smell is a huge technical debt which affects program readability and comprehension. While the presence of design smells does not impede the operation of the software system, it can affect software development, sustainability, and increase the likelihood of software failure \cite{turkistani2019reducing}. Therefore, mitigating design smells in the source code is imperative for enhancing overall software quality. In the ever-evolving landscape of software development, mobile and desktop applications emerge as predominant software systems. Notably, mobile applications undergo more frequent updates compared to their desktop counterparts \cite{mcilroy2016fresh}. Nevertheless, the need for enhanced software maintenance is important for improving software quality irrespective of the underlying ecosystems. 

Responsibility (role) stereotypes indicate generic responsibilities that classes play in the design of a software system. These roles include; Coordinator (CO), Structurer (ST), Controller (CT), Information Holder (IH), Interfacer (IT), and Service Provider (SP) as initially classified by \textcite{wirfs2006characterizing}. In this paradigm, classes are characterized according to the type of responsibility they play in the software design. Knowledge about role-stereotypes has proven very helpful in various tasks of software development and maintenance, such as program understanding, program summarization and quality assurance. Role-stereotypes have also been used in creating better layouts of class diagrams which results in improved program comprehension \cite{genero2008does,nurwidyantoro2019automated, sharif2009effect,ho2022role}.  Besides this, enhancing source code with role-stereotype information helps improve feature location in source code \cite{alhindawi2013improving}. The benefits of role-stereotypes can be observed as an enrichment of reverse engineering, from the perspective of under-covering software design choices \cite{nurwidyantoro2019automated, ho2022role}.

Although the concepts of role-stereotypes and design smells are widely divergent, both are significant contributors to the design and maintenance of software systems. Several studies related to design smells focus on developing detection methods and tools to improve software quality \cite{barbez2020machine,kaur2018detecting, liu2018deep}. On the other hand, studies related to role-stereotypes focuses on classification \cite{dragan2010automatic,nurwidyantoro2019automated} and application of role-stereotypes for example in UML class diagram design \cite{kuzniarz2004empirical}. The substantial interest in enhancing software design and maintainability within the software engineering industry reinforces the importance of understanding the correlation between design smells and role-stereotypes.

 Building this empirical knowledge can provide the following significance; (1) \textit{software maintainability:} understanding the correlation between design smells and role stereotypes can help in identifying patterns that impact software maintainability. This knowledge aids developers in proactively addressing design issues to prevent long-term software degradation, (2) \textit{improving code quality:} identifying and mitigating design smells contributes to overall code quality improvement. Developers can prioritize and address design issues that are more likely to impact the specific role stereotypes, (3) \textit{efficient code reviews:} guide code reviews. Developers can focus on areas of the codebase that are more likely to be associated with specific roles, leading to more efficient and targeted code inspections, (4) \textit{guided code refactoring:} knowing the impact of design smells on role stereotypes provides valuable guidance for refactoring efforts. Developers can prioritize refactoring activities based on the roles affected, ensuring that improvements align with the intended behavior and responsibilities of each software component, (5) \textit{early  defect detection:} this is particularly important in the software development life cycle, where addressing issues early on is more cost-effective than dealing with them in later stages and (6) \textit{role-specific metrics:} the empirical knowledge gained in this area can contribute to the development of role-specific metrics for assessing the health and quality of software components.

Despite the aforementioned significance, the relationship between design smells and role-stereotypes remains unclear due to  the structural complexity of source code across diverse software application ecosystems. To this end, we present an exploratory analysis based on statistical and machine learning methods to understand the relation between design smells and role-stereotypes. Machine learning techniques are able to discover local patterns and make inferences from complex feature representations in a given dataset \cite{awad2015machine}. Recently, unsupervised learning methods have been leveraged to study design smells in static code analysis and have received commendable results \cite{caram2019machine,imran2019design,palomba2017investigating,tahmid2016understanding}. Our study is based on 11,350 Java classes of 30 open source Java-based projects mined from GitHub. To the best of our knowledge, this is the first study on establishing intrinsic relationship between design smells and role stereotypes.
As such, we hope to open up the possibility of enhancing design smell metrics with role-stereotype properties as previously observed \cite{nurwidyantoro2019automated}. The main contributions of the paper are:
\begin{itemize}
	\item [(1)] The study provides a step-by-step approach to build a fine-grained dataset comprising of a combination of design smells and role-stereotype classification data. As a result, we publish a sizable dataset of 11,350 Java classes which can serve as a resource for other researchers.
	\item [(2)] Using the perspective of role-stereotypes, we presents a comparison of the occurrence of design smells in desktop versus mobile applications. We show the relevance of role-stereotype information in dealing with design smells.
	\item [(3)]The study provides a clustering approach to analyze the groupings of role-stereotypes that are prone to similar categories of design smells.
	\item [(4)] Finally, the paper provides insights to software developers, designers and researchers on previously concealed behavior and relationships between design smells and role-stereotypes.
\end{itemize}

The rest of this paper proceeds as follows: In Section~\ref{sec:terminologies}, we present the background of the study, derive terminologies and provide a concrete example to motivate the need for the study. Section~\ref{sec:relatedwork} outlines related work and identifies study gaps. In Section~\ref{sec:method}, we provide a comprehensive explanation of the research questions, tools, methods, and analysis performed. The results of our analysis are presented in Section~\ref{sec:result}. Section~\ref{sec:ds_imp} focuses on discussions and implications of the study. The threats to validity are discussed in Section~\ref{sec:threats}. Finally, Section~\ref{sec:conclusion} concludes the paper and provides direction for future work.
\section{BACKGROUND}\label{sec:terminologies}
In this section, we derive important terminologies used throughout the study and provide a concrete example illustrating the intrinsic relationship between design smells and role-stereotypes, which can be observed through code inspection.  The objective of this illustration is to validate the co-occurrence of design smells and role stereotypes in actively maintained codebases, highlighting the necessity for deeper exploration into their co-occurrence dynamics. As previously noted, while role stereotypes and design smells are loosely distinct concepts, both exert significant influence on the design and maintenance of software systems.
\subsection{Terminology}
\begin{itemize}[leftmargin=*]
	\item \textbf{Role stereotypes:} Denote generic responsibilities that classes undertake in the design of a software system. We employ terms such as ``class role'', ``class responsibility'', and ``class role stereotypes'' interchangeably. It is important to acknowledge that the formal definition is derived from a software design standpoint. However, in this paper, we examine role stereotypes from the implementation perspective.
	
	\item \textbf{Design smells:} Indication of poor design and implementation choices, leading to problematic source code structures. In the paper, the term ``antipattern'', ``bad smell'', ``code smell'' has the same meaning as design smell.
	\item \textbf{Class path:} Refers to the fully qualified class name or file name of a Java class. Following \textcite{runeson2012case} recommendation, we use Java classes as our unit of analysis.
	\item \textbf{Regex:}  Refer to regular expression, used to extract class path from the output of design smell detection.
	\item \textbf{Co-occurrence:} Refers to simultaneous presence of two or more design smells within a single role stereotype. For example; if ``LongParameterList`` and ``ComplexClass'' are detected together in Information Holder role stereotype, then we say  ``LongParameterList`` and ``ComplexClass'' co-occur.
	\item \textbf{Cluster}: In this study, a cluuster refers to group of design smells or role-stereotypes with shared characteristics.
\end{itemize}
\subsection{Concrete Example}
To put the problem into context, let us consider a practical example involving code refactoring as a use case. In Listing \ref{lst:rscode}, we examine the {\tt ImapStoreSettings.java} class within the \textbf{K9 Mail app}\footnote{\url{https://github.com/thunderbird/thunderbird-android}}. This class encapsulates IMAP store settings, offering methods for accessing, manipulating, and retrieving additional settings in key-value pairs. We categorize this class under the stereotype of a {\tt Information Holder} due to its primary function of managing key-value pairs, a characteristic central to the {\tt Information Holder} role as described. It is evident that the code is impacted by the {\tt LongParameterList} design smell, highlighted in lines {\tt 10-12}. In Listing \ref{lst:dscode}, we present a sample output of the {\tt LongParameterList} design smell detected in the K9 Mail project using the Software Architectural Defects (SAD) tool, with line {\tt 19} showcasing the {\tt ImageStoreSettings.java} class.

Assuming a developer aims to refactor the code at lines {\tt 10-12} in Listing \ref{lst:rscode}, it becomes essential to understand not only the design smell and the class role stereotype but also how they are interrelated. For instance, the developer might opt to address the issue by implementing the \textit{Builder} design pattern. This involves creating a \textit{Builder} class responsible for constructing an {\tt ImapStoreSettings} object, allowing clients to set specific attributes before building the final object. However, this approach might be more suitable for a {\tt Controller} class role, as the \textit{Builder} pattern is better suited for object construction. Since the {\tt Information Holder} class focuses on data transfer, a more appropriate refactoring strategy would involve implementing a structural design pattern, such as the Data Transfer Object (DTO) pattern.
 
	\begin{lstlisting}[language={java}, mathescape, framexleftmargin=3em, frame=trlb,caption={\footnotesize Code snippet of  {\tt ImapStoreStettings.java} IH class of K9 Mail project. Line {\tt 10-12} highlighted indicate a {\tt LongParameterList} design smell\label{lst:rscode}}, numbers=left, escapechar=!, tabsize=2, basicstyle=\ttfamily\footnotesize]
	/**
	* This class is used to store the decoded contents of an ImapStore URI.
	*
	* @see ImapStore#decodeUri(String)
	*/
	public class ImapStoreSettings extends ServerSettings {
 
		...
		
		!\hl{protected ImapStoreSettings(String host, int port, ConnectionSecurity connectionSecurity,}!
		!\hl{AuthType authenticationType, String username, String password, String clientCertificateAlias,}!
			!\hl{boolean autodetectNamespace, String pathPrefix)\{}!
			
			super(Type.IMAP, host, port, connectionSecurity, authenticationType, username,
			password, clientCertificateAlias);
			
			this.autoDetectNamespace = autodetectNamespace;
			this.pathPrefix = pathPrefix;
		}
		
		...
	}
\end{lstlisting}

\begin{lstlisting}[language={ini}, mathescape, framexleftmargin=3em, frame=trlb,caption={\footnotesize Example {\tt LongParameterList} design smell detected from K9 Mail project using SAD tool. Line {\tt 19} shows {\tt ImageStoreSettings.java} class\label{lst:dscode}}, numbers=left, escapechar=!,tabsize=2, basicstyle=\ttfamily\footnotesize]
	# Results of the detection 
 
	...
 
	# ------>LongParameterList num: 17
	
	17.100.Name = LongParameterList
	
	#LongParameterListClass
	17.100.LongParameterListClass-0 = k9mail.src.main.java.com.fsck.k9.activity.compose.RecipientPresenter
	17.100.LongParameterListClass-0.NOParam-0 = 9.0
	17.100.LongParameterListClass-0.NOParam_MaxBound-0 = {NOParam_MaxBound=6.0}
	
	# ------>LongParameterList num: 18
	
	18.100.Name = LongParameterList
 
	#LongParameterListClass
	!\hl{18.100.LongParameterListClass-0 = k9mail-library.src.main.java.com.fsck.k9.mail.store.imap.ImapStoreSettings}!
	18.100.LongParameterListClass-0.NOParam-0 = 9.0
	18.100.LongParameterListClass-0.NOParam_MaxBound-0 = {NOParam_MaxBound=6.0}
	
	...
\end{lstlisting}
Form the above code snippets, we can ask several questions. For example, are Information Holders more prone to {\tt LongParameterList} than other role-stereotypes? How often does {\tt LongParameterList} occur? Is the occurrence influenced by the type of application - desktop or mobile app? What design choices should be considered for Information Holder role-stereotypes to mitigate occurrence of {\tt LongParameterList}? What refactoring opportunities are possible and how can they be applied? To answer those questions and many other related ones, it is important to build empirical knowledge on the relationship between design smells and role-stereotypes. 

\section{RELATED WORK}\label{sec:relatedwork}

This section explores the existing body of work surrounding design smells and role stereotypes with the aim of uncovering variations in their relationship. As mentioned in the introduction section, the presence of design smells can hinder the evolution and performance of software systems. Similarly, role stereotypes provide a lens through which the responsibilities and interactions of software components can be understood and optimized. Despite their importance, the interplay between these two domains has not been thoroughly explored in the current body of literature. By examining various detection methodologies and the implications of role stereotypes on software design, this section sets the stage for a deeper understanding of how design smells and role stereotypes influence software systems.
\subsection{Design Smells}

Design smells incorporate low-level or local issues in the source code, commonly known as code smells \cite{vaucher2009tracking}. These code smells are indicative of potential defects within the source code that may compromise software quality and evolution \cite{soh2016code, oizumi2018identification, abuhassan2021software}. Fowler famously cataloged 22 code smells and proposed corresponding refactoring techniques to address these issues \cite{1999:RID:311424}. Although code smells do not necessarily impede software operation, their presence can lead to a range of complications, such as diminishing the software’s sustainability and increasing its likelihood of failure \cite{turkistani2019reducing}. Thus, addressing design and code smells is paramount for ensuring software scalability and maintainability. Refactoring, which involves modifying the internal structure of software code without altering its external behavior, is widely recommended. This process not only enhances software maintainability but also leads to a more coherent internal architecture \cite{turkistani2019reducing}. Design smells often encompass broader aspects of software structure and impact; therefore, refactoring a design smell typically requires modifications across multiple classes \cite{sharma2020empirical}. Several researchers have investigated various methodologies for identifying code smells within source code, each utilizing distinct detection techniques. These methodologies are discussed below:
\begin{itemize}[leftmargin=*]
    \item \textbf{Metric-based Detection:} This method employs quantitative measures to identify potential issues in the source code. Metrics can include complexity scores, coupling between objects, or lines of code, which help in pinpointing areas that may harbor code smells. \cite{bafandeh2020bad} used a metric-based design smell detection approach to detect bad code smells. This was achieved through cross-referencing each instance of the detected bad smell with the corresponding refactoring technique. \cite{kovavcevic2024automatic} introduced a data-driven method for setting threshold values for object-oriented metrics in code smell detection, using statistical properties of metrics from 74 systems in the Qualitas Corpus. This transparent, repeatable approach allows for the calibration of detection rules to be more contextually appropriate, reducing the occurrence of false positives and negatives in identifying design flaws in code.
    \item \textbf{Machine Learning Techniques:} Machine learning offers a dynamic way to detect code smells by learning from examples. This method can adapt to different coding styles and environments, potentially providing more accurate detections over time as it learns from new data.  Kaur \& Singh \cite{kaur2018detecting} presented a supervised-based machine learning algorithm for detecting software code smells from design patterns. The authors considered four (4) design patterns and five (5) code smells. Using J48 decision tree classifier, it was observed that there is a relationship between some design patterns with specific code smells. Additionally, several authors have applied deep learning and other traditional machine learning models to detect Feature Envy design smells \cite{liu2018deep, liu2023deep, vskipina2024automatic, alazba2024cort}. \cite{barbez2020machine} extended the work in \cite{liu2018deep} using an ensemble model called Smart Aggregation of Anti-patterns Detectors (SMAD). The authors reported that SMAD outperformed other ensemble methods in detecting God Class and Feature Envy antipatterns in eight Java projects. Despite the promise of machine learning-based smell detection, its effectiveness is constrained by the need for large training datasets, which remains a significant challenge.
    
    \item \textbf{Optimization-based Detection:} Optimization techniques are used to address code smells by framing the detection as an optimization problem. This involves finding the best solution that minimizes or maximizes a particular objective, such as reducing the complexity of the code. Saranya et al. \cite{saranya2018model} implemented an optimization-based approach for detecting model-level code smells, utilizing a combination of Euclidean distance-based Genetic Algorithm and Particle Swarm Optimization (EGAPSO). This study was motivated by the limitations inherent in metric-based code smell detection methods.
    \item \textbf{LLM-based Detection:} Large Language Models (LLMs), like GPT and others, are increasingly being used in software engineering tasks. These models leverage their vast knowledge base and understanding of programming semantics to identify subtle issues in the code that traditional methods might miss. Liu et al. \cite{liu2024prompt} conducted an earlier study on detecting code smells using the LLM approach. In this study, the authors proposed \textit{PromptSmell}, a novel approach based on prompt learning for detecting multi-label code smell. The authors evaluated the effectiveness of this approach and the experimental results demonstrated that \textit{PromptSmell} obtains an improvement of 11.17\% in precision and 7.4\% in F1-score compared to existing approaches.
\end{itemize}

\subsection{Role-Stereotypes}
Role-stereotypes indicate generic responsibilities that classes play in the design of a system. \cite{wirfs2006characterizing} proposed six (6) generic categories summarized below:
\begin{itemize}
    \item {\it Coordinator}: An object that does not make many decisions but, in a rote or mechanical way, delegates work to other objects.
    \item {\it Structurer}:  An object that maintains relationships between objects and information about those relationships. 
    \item {\it Controller}: An object designed to make decisions and control a complex task.
    \item {\it Information Holder}: An object designed to know certain information and provide that information to other objects. 
    \item {\it Interfacer}: An object that transforms information or requests between distinct parts of a system.
    \item {\it Service Provider}: An object that performs specific work and offers services to others on demand. 
\end{itemize}
This classification is intended to convey an important part of the design intention of a class. The benefits of role stereotypes are observed in software development and maintenance tasks, such as program comprehension, program summarization, quality assurance and in the creation of layouts for class diagrams \cite{genero2008does, nurwidyantoro2019automated, sharif2009effect}.
Based on the benefits that role-stereotypes offer to software design, several researchers have explored the classification of class role-stereotypes in source code. Earlier work by \textcite{dragan2010automatic} focused on automatic classification of a class's stereotype for C++ source code. The proposed method consists of rules based on both the stereotype and category distributions of the class signature. \textcite{moreno2012jstereocode} extended Dragan's work to Java source code. 
They developed ``{\tt JStereoCode}'', a tool that automatically identifies the stereotypes of methods and classes in Java systems. The studies in \cite{dragan2010automatic} and \cite{moreno2012jstereocode} are based on expert-designed decision rules that are applied to the syntactical characteristics of the class source code. \textcite{nurwidyantoro2019automated} provides a more recent study in this direction by applying machine learning models for improved automated classification of role-stereotypes in Java classes. Source code classes were classified in one of the six role-stereotypes taxonomy proposed by  \textcite{wirfs2006characterizing}. The result shows that the Random Forest algorithm enhanced by SMOTE resampling yielded the best performance compared to Multinomial Naive Bayes (MNB) and Support Vector Machine (SVM) models.

There are notable trends of research on the use of role stereotypes to enhance software comphresion. Research by \textcite{ahmed2020classification} used machine learning to classify role stereotypes in UML class diagrams, enhancing early software design understanding. By labeling 391 classes from 15 open-source projects, they created a ground truth and found that J48 classifier is most effective on raw data, while Random Forest model excelled on a balanced dataset via SMOTE. This method supports software quality assessment and design summarization, offering early analysis tools for developers.
\textcite{ho2020interactive} developed ``{\tt RoleViz}'', an interactive visualization tool that enriches traditional software visualization by incorporating role stereotypes. A user study involving 16 developers compared ``{\tt RoleViz}'' with Softagram, a commercial software architecture comprehension tool, across cognitive load, usability, and understanding of a large open-source system. The results showed that RoleViz significantly enhanced participants' understanding of the system without increasing cognitive load, with better usability scores, and six participants specifically noted that visualizing roles facilitated their comprehension tasks.
Another interesting study by \textcite{gokmen2021stereotypes} proposes using stereotypes as design patterns in serious games to simplify the understanding of object-oriented software. It emphasizes that incorporating software knowledge into game mechanics not only enhances learning and engagement but also facilitates the application of these games across various software systems. The authors recommends the development of automatic tools for identifying stereotypes to expedite the game design process, making serious games a more effective tool for software comprehension for both developers and non-experts.

\subsection{Research Gaps}
In this section, we discuss related work done to understand the relationship between design smells and role-stereotypes. We also try to unveil some gaps in the current body of literature to justify the need for this study. It is important to note that, to the best of our knowledge, no formal in-depth study has been done so far to assess the relationship between design smells and role-stereotypes. Most research attention from both academia and software industry has been directed towards developing various detection and classification methods and tools for design smell and role-stereotype respectively. Some authors have tried to study the relationship between design smells and design patterns \cite{jaafar2014analysing, walter2016relationship, sousa2018systematic, santos2022building} but not role-stereotypes.

\noindent \textcite{jaafar2014analysing} analyzed the static relationship between anti-patterns and design patterns. They studied anti-patterns dependencies with other classes within particular design patterns to understand how developers can maintain programs containing those anti-patterns. The study was performed on 1,191 to 3,325 classes of three (3) Java projects (\textit{ArgoUML, JFreeChart, and XercesJ}). It was observed that there is a temporary relationship between anti-pattern and design patterns. However, this study strictly focused on dependencies between anti-patterns and design patterns but not role-stereotypes. Besides, only particular sets of anti-patterns and design patterns were used yet different anti-patterns and design patterns could potentially influence the outcome. The work in \cite{jaafar2014analysing} did not consider the type of systems (desktop, mobile or web-based) and how system type can impact the dependencies between anti-patterns and design patterns.

\noindent \textcite{walter2016relationship} conducted a study related to that of \cite{jaafar2014analysing} to understand how the presence of design patterns impacts the existence of code smells. Based on an exploratory approach, 9 design patterns and 7 code smells were analyzed from two sizeable open-source Java projects. Their findings indicated that classes which participate in design patterns appear to display code smells less frequently than other classes. The observed effect was stronger for some patterns (e.g., Singleton, State-Strategy) and weaker for others (e.g., Composite). However, this study does consider whether role-stereotypes could influence the association of design patterns with code smells. The type of application was also not considered in this study. The impact of design patterns on the presence of code smells might differ between desktop, mobile and web applications.

\noindent \textcite{mannan2016understanding} focused mainly on code smells which tends to affect readability and simplicity. The task of refactoring in this case involves renaming or extracting to methods. Design smells, on the other hand, tend to be more subtle. They usually affect maintainability and flexibility. Next, their study did not look at the general co-occurrence of code smells and how these co-occurrences varies across desktop and android application. In addition, this study did not look at role-stereotypes. \textcite{palomba2017investigating} studied the co-occurrence of different types of smells on the same code component. It was observed that some code smells frequently co-occur and the occurrence of class-level seems to originate from method-level code smells. The projects selected in this study comprises of mixture of Java desktop applications and libraries, whose internal implementation slightly differs. Furthermore, the study did not look at the association of role-stereotypes and co-occurrence of code smells.

In summary, while there has been some research on understanding design smells and role-stereotypes, there remains a significant gap in exploring the non-inherent relationships between them. Additionally, how these relationships are influenced by the type of application, whether mobile or desktop, has not been thoroughly investigated.

\section{STUDY DESIGN}\label{sec:method}
Overall, we are interested in building empirical knowledge on the relationship between design smells and role stereotypes. To this end, we outline three research questions detailed in subsection~\ref{sec:researchquestion}, followed by data collection and analysis steps.
\subsection{Research Questions}\label{sec:researchquestion}
Specifically, our study aims at answering the following research questions: 
\begin{itemize}[leftmargin=*]
	\item \textbf{RQ1:} \textit{How do design smells vary across desktop and mobile applications?} The aim is to investigate and understand the differences in the occurrence and distribution of design smells in software systems developed for desktop applications compared to mobile applications. We are also interested in the co-occurrence of these smells. This will give us better understanding of the nature and characteristics of design smells across software ecosystems. \textcite{martins2020code} noted the need for further research on the impact of code smells co-occurrences on internal quality attributes.
	\item \textbf{RQ2:} \textit{How do design smells relate with different types of role-stereotypes?} In this research question, the focus is to explore the associations and dependencies between design smells and various role-stereotypes within the context of software systems. For example: Do `God-classes’ appear (significantly) more often with \textit{Controller} role-stereotype than with \textit{Service Provider} role-stereotypes? This will provide valuable insights that can guide software developers in making informed decisions during the design and refactoring activities.
	\item \textbf{RQ3:} \textit{Does the type of application (desktop or mobile) influence the relation between design smells and role-stereotypes?} The focus is on understanding whether the nature of the application platform i.e. desktop or mobile, has an impact on the interplay between design smells and role-stereotypes within software systems. We want to understand the relationship and variations between design smells and role-stereotypes based on the software ecosystem?
\end{itemize} 

\subsection{Data Collection}
\begin{figure}[h]
	\centering
	\includegraphics[scale=0.7]{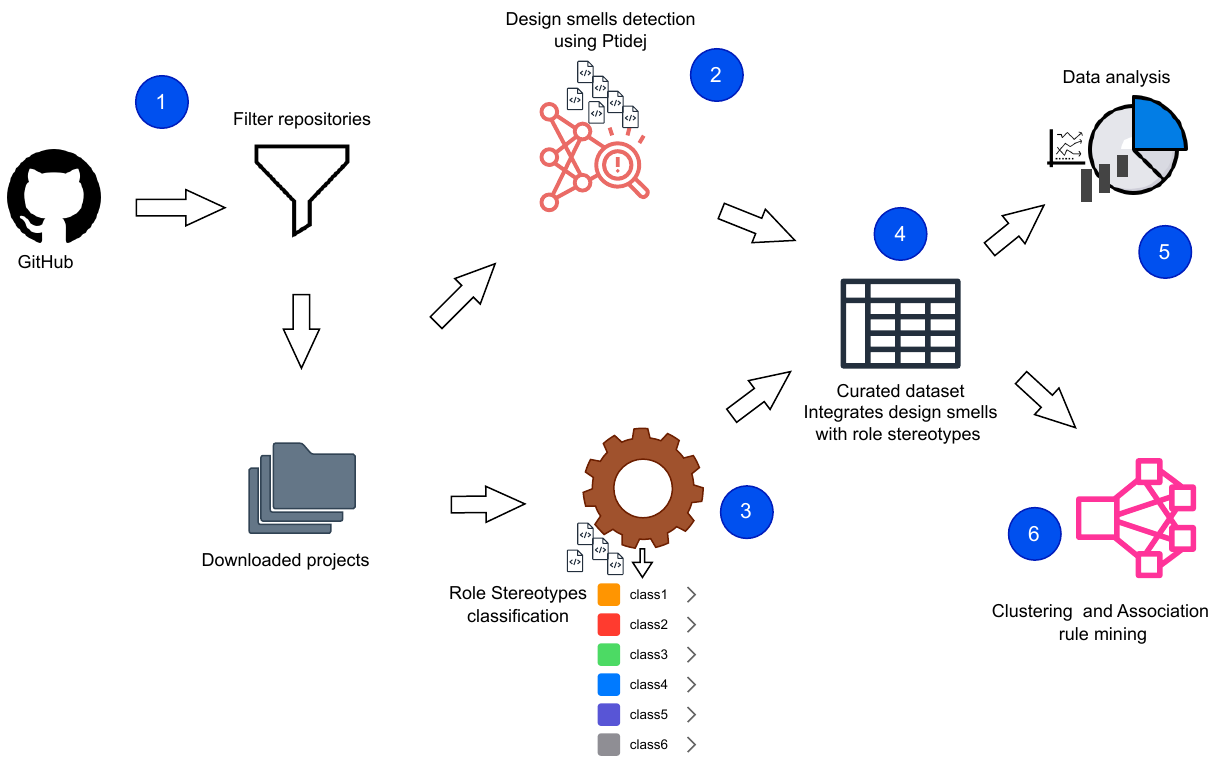}
	\caption[Methodology Pipeline]{The research methodology pipeline. Step 1: data collection; Step 2: design smell detection; Step 3: role-stereotype classification; Step 4: integrate design smell and role-stereotype data to form a fine-grained dataset; Step 5 \& Step 6: data analysis and clustering.}
	\label{fig:conceptualfw}
\end{figure}

As shown in Figure \ref{fig:conceptualfw}, \textit{Step 1} describes the data collection and selection strategy. \textit{Step 2} focuses on the detection and preprocessing design smells from the selected projects. \textit{In Step 3}, we preprocess and classify role stereotypes into 6 categories as previously discussed. In \textit{Step 4}, we systematically integrate design smells and role-stereotype data to produce a fine-grained dataset. Using the fine-grained dataset, we perform analysis in \textit{Step 5} and data clustering in \textit{Step 6}  to answer the aforementioned research questions.

\subsubsection*{Step 1: Project Selection}

We built a dataset based on 30 active open source Java projects mined from GitHub containing a total of 11,350 number of classes (NOC) and 2,224,258 LOC ($\sim$2,224 KLOC). The projects selected in this study are commonly studied by previous researchers and predominantly utilize Java as their primary programming language \cite{nurwidyantoro2019automated, ho2022role}. The decision to use Java for this study is primarily driven by two key factors. First, Java is compatible with the design smell detection tool we have chosen, as detailed in Section~\ref{sec:terminologies}. This tool is specifically tailored for Java, making it the most effective language for our analysis. Second, Java's widespread adoption in a wide variety of object-oriented, open-source projects across multiple code repositories makes it an ideal candidate for assessing a diverse range of design smell metrics. According to \textcite{mannan2016understanding}, Java has more design smell detectors than other languages. Furthermore, Java language is known to support the operation of billions of devices globally. Consequently, the chosen projects are cross-platform compatible (i.e., they can operate on Windows, macOS and Linux operation systems) and depend on Java core libraries for the design of their logical layers. Given these general characteristics, we curated a set of 30 projects from GitHub, provided that the project had at least two contributors and the latest commit was not more than one year. This was to ensure that the projects were still actively maintained (not ``toy'' projects) and modest in size. We selected GitHub because of its popularity as the largest social coding platform, hosting the development history of millions of collaborative software repositories and  offering diverse source code metadata~\cite{decan2022use}.
The selection of Android-based mobile projects in this study allows for a valid comparison with desktop projects since Android source code is often written in Java or Kotlin. Furthermore, Android stands as the leading mobile operating system, commanding a market share of up to 74.6\%\footnote{\url{https://gs.statcounter.com/os-market-share/mobile/worldwide}}. It also consists of a large community of developers delivering billions of apps on mobile software ecosystem. The selected projects are grouped as either ``desktop'' or ``mobile'' software project. Tables \ref{table:selected_projects-desktop} and \ref{table:selected_projects-mobile} describes the characteristics of projects selected for this study including their domain, version LOC and NOC. 

\begin{table*}[h]
	\centering 
	\caption{Characteristics of desktop projects analyzed in this study (LOC: Lines of Code; NOC: Number of Classes). } 
	\begin{tabular*}{\textwidth}{@{\extracolsep\fill}lllllll@{\extracolsep\fill}}%
		\toprule
		\textbf{No.} & \textbf{Project} & \textbf{Description}  & \textbf{Version} & \textbf{LOC} & \textbf{NOC} \\ 
		\midrule
		1 & SweetHome3d & Interior Design Tool & 5.6 & 104,059 & 546 \\ 
		2 & Mars Simulation & Object Modeling & 3.1.0 & 255,459 & 1109 \\ 
		3 & ArgoUML & UML Modeling Tool & 0.35.1 & 177,372 & 1236 \\ 
		4 & JEdit & Text Editor & 5.5.0 & 124,164 & 577 \\ 
		5 & GanttProject & Project Scheduling  and Management & 2.9.11 & 66,709 & 671 \\ 
		6 & GoGreen & Carbon Footprint Tracker & 0.1.3 & 4,196 & 60 \\  
		7 & LiveChatServer & Live Chat Tool & 1.4 & 2,796 & 23 \\ 
		8 & Checkstyle &  Java Source Code Validator  & 8.39 & 238,255 & 1008 \\ 
		9 & Keystore-explorer & Secret Key Management & 5.4.4 & 55,875 & 400 \\ 
		10 & Angry IP Scanner & Network Scanner & 3.7.3 & 12,701 & 159 \\
		11 & JetUML & UML Modeling Tool & 3.1 & 24,965 & 173 \\ 
		12 & JPass & Password Manager &  0.1.20 & 3,442 & 38 \\ 
		13 & LogFX & Log Reader & 0.9.1 & 4,582 & 44 \\ 
		14 & PGP Tool & Easy PGP Decryption  and encryption  & 0.5.9.2 & 18,673 & 226 \\ 
		15 & Freemind & Mind-mapping Tool & 1.0.0 & 67,287 & 370 \\  
		\bottomrule
	\end{tabular*}
	\label{table:selected_projects-desktop} 
\end{table*}

\begin{table*}[h]
	\caption{Characteristics of mobile projects analyzed in this study (LOC: Lines of Code; NOC: Number of Classes). } 
	\centering 
	\begin{tabular*}{\textwidth}{@{\extracolsep\fill}lllllll@{\extracolsep\fill}}%
		\toprule
		\textbf{No.} & \textbf{Project} & \textbf{Description}  & \textbf{Version} & \textbf{LOC} & \textbf{NOC} \\ 
		\bottomrule
		1 & K9 Mail & Email Client App & 5.600 & 93,540 & 779 \\ 
		2 & Bitcoin Wallet & Bitcoin Payment App & 6.31 & 18,079 & 222 \\ 
		3 &  KeepassDroid & Password Manager & 2.5.9 & 17,916 & 211 \\ 
		4 & Opentrip Planner &  Trip Planning and Navigation & 2.1.5  & 9,760 & 53 \\ 
		5 & Tweet Lanes & Twitter Client App & 1.4.1 &  541,694 & 130 \\ 
		6 & Signal Android & Messaging App & 4.69.5  & 25,886 & 1332 \\ 
		7 & Telegram & Messaging App & 4.1.1  & 166,731 & 679 \\ 
		8 & Materialistic & News Reader App & 3.3  & 21,919 & 131 \\ 
		9 & Telecine & Full Resolution Video Recoder  & 1.6.2 & 1,410 & 23 \\ 
		10 & AmazeFileManager & File Manager &  3.5.2  & 46,135 & 265 \\ 
		11 & Omni-Notes & Note Taking App & 6.0.5 &  15,932 & 159 \\ 
		12 & AntennaPod & Podcast Manager & 2.1.1 & 53,769 & 387 \\ 
		13 & GnuCash & Expense Tracker & 2.4.1 & 27,837 & 147 \\ 
		14 & Timber & Music Player & 1.8 & 20,562 & 163 \\ 
		15 & SeeWeather  & Weather App & 2.03 & 2,553 & 29 \\ 
		\bottomrule
	\end{tabular*}
	\label{table:selected_projects-mobile} 
\end{table*}

\subsubsection*{Step 2: Design smells detection}



To detect design smells in source code, several tools are available, including SonarQube, PMD, JDeodorant, DERCOR, FindBugs, and DesigniteJava, among others. However, for this study, we utilized the SAD component of the Pattern Trace Identification, Detection, and Enhancement in Java (Ptidej) tool suite \footnote{\url{https://github.com/ptidejteam/v5.2}}. Ptidej is a Java-based reverse engineering tool suite equipped with various algorithms for identifying idioms, micro-patterns, design patterns, and design defects \cite{gueheneuc2007ptidej}. As an open-source platform, Ptidej is particularly tailored for detecting design patterns and anti-patterns in Java code, making it highly beneficial for academic research and quality assurance in software development. It is recognized as one of the most popular static code analysis tools in academic research \cite{10.1007/978-3-030-14687-0_15, 10.1145/3345629.3345630} and has been employed in over 200 research projects \footnote{\url{http://www.ptidej.net/publications/}}. Table \ref{table:ds_types} outlines the 18 different types of design smells detectable by Ptidej. Using Java classes as the unit of analysis, SAD detects occurrence of design smell using a set of rule cards. For example; ``LongParameterList'' can be detect using the following rule card:
\begin{verbatim}
	RULE_CARD : LongParameterList { 
		RULE : LongParameterListClass { (METRIC: NOParam, VERY_HIGH, 6) } ; 
	};
\end{verbatim}
Using the rule card provided, the SAD tool performs a scan on Java classes, parsing their methods' signatures. For each method, it checks the number of parameter lists against the threshold specified by the \textit{NOParam} metric (which is set to less than 6). If the count of parameters exceeds or equals 6 (\textit{NOParam} >= 6), the tool identifies a ``LongParameterList'' occurrence within that class and saves the result in a ``.ini'' file. A sample output of this process is shown in Listing \ref{lst:dscode}.

\begin{table*}
\fontsize{8pt}{8pt}\selectfont
	\centering 
	\caption{Description of the design smells detected using the Ptidej tool.} 
	\begin{tabular*}{\textwidth}{@{\extracolsep\fill}lll@{\extracolsep\fill}}%
			\toprule
			\textbf{No.} & \textbf{Design Smell} & \textbf{Description} \\ 
			\midrule
			1 & AntiSingleton & Provides mutable class variables, which consequently could be used as global variables. \\ 
			2 & BaseClassKnowsDerivedClass & A class that has many subclasses without being abstract.\\ 
		3 &BaseClassShouldBeAbstract &  A class that has many subclasses without being abstract. \\ 
			4 &Blob & Large class declares many fields and methods with a low cohesion. \\
		5 &ClassDataShouldBePrivate & A class exposing its fields, violating the principle of data hiding.\\ 
		6 &	ComplexClass & A class having at least one method having a high cyclomatic complexity. \\ 
		7 & FunctionalDecomposition & A main class with a procedural name in which inheritance and polymorphism are scarcely used.\\ 
		8 &	LargeClass  & A class that has grown too large in term of Lines of Code.\\ 
		9 &	LazyClass & A class having very small dimension, few methods and low complexity. \\ 
		10 &	LongMethod & A method that is unduly long in terms of lines of code.\\ 
		11 &	LongParameterList & A method having a long list of parameters, some of which avoidable.\\
		12 &	ManyFieldAttributesButNotComplex & Declares many attributes but which is not complex. Data class holding values without providing behaviour.\\
		13 &	MessageChains  & A long chain of method invocations performed to implement a class functionality. \\ 
		14 &	RefusedParentBequest & A class redefining most of the inherited methods, thus signaling a wrong hierarchy. \\ 
		15 &	SpaghettiCode  & A class implementing complex methods interacting between them, with no parameters, using global variables.\\ 
		16 &	SpeculativeGenerality & A class declared as abstract having very few children classes using its methods. \\ 
		17 &	SwissArmyKnife  & Complex class that offers a high number of services, such as implementing a high number of interfaces.\\
		18 &	TraditionBreaker & A class that inherits from a large parent class but that provides little behaviour and without subclasses.\\
			\bottomrule
		\end{tabular*}
		\label{table:ds_types} 
	\end{table*}
 \noindent Design smells are detected and stored in ``{\tt .ini}'' files. The file names are tagged with a specific design smell type. For example, in the K9 mail project, {\tt LongParameterList} design smell is stored as {\tt DetectionResults in K9 for LongParameterList.ini}. Our goal is to extract class names and the corresponding design smell detected in that class. Listing~\ref{lst:dscode} shows a sample content of design smell detection file in its raw format. Based on the structure of the \textit{``.ini''}, we derived custom regular expressions to extract class names and their corresponding design smells. In the case of our example in Listing~\ref{lst:dscode}, we use this regex: {\tt k9mail[a-zA-Z0-9.-]+}. For each design smell, we count the number of its occurrence in a given class, otherwise, the value {\tt 0} is assigned \cite{ogenrwot, ogenrwot2021integration, ogenrwot2021using}. The summary of this process is illustrated as pseudo-code in Algorithm \ref{al:process_ds} and the actual implementation can be obtain from our replication package \cite{ogenrwot_2024_10775827}. Table \ref{table:ds_processed} shows a sample output of processed design smells data.

\begin{algorithm}
	\caption{\enskip Peusocode for Extracting Design Smells}\label{alg1}
	\begin{algorithmic}[1]
		\Procedure{ExtractDesignSmells}{$iniFilePath$}
		\State $result \leftarrow \{ \}$
        
		\For {$file \in iniFilePath$}
		\For{$smell \in smellList$}
		\State $fileContent \leftarrow readFileContent(file)$
		\State $matches \leftarrow findDesignSmell(fileContent, smell)$
        
		\If{|matches| $\neq$ 0}
		\State $columnValue \leftarrow columnValue + 1$
		\Else
		\State $columnValue \leftarrow 0$
		\EndIf
        
		\State $rowValue \leftarrow file$
		\State $result.add(rowValue, columnValue)$
		\EndFor
		\EndFor
        
		\State \Return $result$
		\EndProcedure
	\end{algorithmic}
	\label{al:process_ds}
\end{algorithm}

		\begin{table*}[h]
  \fontsize{9pt}{9pt}\selectfont
		\centering 
		\caption{Sample output of processed design smells dataset. The columns indicate the design smell type and the row indicate Java classes.} 
			\begin{tabular*}{\textwidth}{@{\extracolsep\fill}lllllll@{\extracolsep\fill}}%
			
			\toprule
			\textbf{index}& \textbf{FullClassPath} & \textbf{Classname} & \textbf{Blob} &\textbf{LongMethod} &\textbf{ LazyClass} & \textbf{...}  \\ 
			\midrule 
			1 & k9mail-library.src.main.java.com.fsck.k9.mail.AuthType.java & AuthType & 3 & 1 &  0 & ... \\ 
			2 & k9mail-library.src.main.java.com.fsck.k9.mail.Address.java & Andress & 1 & 0 & 0 & ...  \\
			3 & k9mail-library.src.main.java.com.fsck.k9.mail.Body.java & Body & 0 & 0 & 2 &  ... \\
			4 &k9mail-library.src.main.java.com.fsck.k9.mail.Flag.java & Flag & 1 & 0 & 0 & ... \\ 
			5 & k9mail-library.src.main.java.com.fsck.k9.mail.Folder.java & Folder & 1 & 0 & 0 & ... \\
			6 &k9mail-library.src.main.java.com.fsck.k9.mail.K9MailLib.java & K9MailLib & 0 & 3 & 0 & ... \\
			7 & k9mail-library.src.main.java.com.fsck.k9.mail.Message.java & Message & 0 & 0 & 1 & ... \\
			8 &k9mail-library.src.main.java.com.fsck.k9.mail.Throttle.java & Throttle & 1 & 2 &0 &  ...  \\
			9 & k9mail-library.src.main.java.com.fsck.k9.mail.K9.java & K9 & 4 & 0 & 0 & ... \\ 
			10 & k9mail-library.src.main.java.com.fsck.k9.mail.MailService.java & MailService & 1 & 2 & 0 & ... \\  
			\bottomrule
		\end{tabular*}
		\label{table:ds_processed} 
	\end{table*}
	
	\subsubsection*{Step 3: Role stereotypes classification}

Following the methodologies established by \textcite{nurwidyantoro2019automated} and \cite{ho2022role}, we employed a set of 23 features for each project, previously identified and validated in their research for classifying role stereotypes. These features serve as the foundation of our unlabeled dataset. The comprehensive description of these features is detailed in the aforementioned studies. Initially, the selected project source code is processed using srcML\footnote{https://www.srcml.org/}, a lightweight, scalable, and robust multi-language parsing tool. This tool converts source code into an Abstract Syntax Tree (AST) represented in XML format. The output from the srcML tool is a list of source code classes in a standardized XML format. Subsequently, we constructed an unlabeled dataset by extracting the 23 features from each project through multiple XPath queries aimed at isolating the features of interest. Finally, the unlabeled data was classified into one of the role-stereotype classes using random forest classifier which obtained the best result when trained and validated on a ground-truth dataset consisting of 773 classes \cite{nurwidyantoro2019automated}. Table~\ref{table:rs_sample} presents a sample of the processed role stereotype data. This methodological approach ensures a rigorous classification process, leveraging both established features and advanced parsing tools to accurately determine role stereotypes within the source code.




	\begin{table*}[h]
        \fontsize{9pt}{9pt}\selectfont
		\centering 
		\caption{Showing sample of processed role-stereotype classification data.} 
		\begin{tabular*}{\textwidth}{@{\extracolsep\fill}lllllll@{\extracolsep\fill}}%

				\toprule
			\textbf{No.}& \textbf{FullClassPath} & \textbf{Classname} & \textbf{loc} &\textbf{numAttr} & \textbf{...} & \textbf{label} \\ 
			\midrule 
			1 & k9mail-library.src.main.java.com.fsck.k9.mail.AuthType.java & AuthType & 33 & 6 & ... & Service Provider\\ 
			2 & k9mail-library.src.main.java.com.fsck.k9.mail.Address.java & Andress & 331 & 4 & ... & Interfacer \\
			3 & k9mail-library.src.main.java.com.fsck.k9.mail.Body.java & Body & 24 & 0 & ... & Interfacer\\
			4 & k9mail-library.src.main.java.com.fsck.k9.mail.Flag.java& Flag & 64 & 15 & ... & Information Holder\\ 
			5 & k9mail-library.src.main.java.com.fsck.k9.mail.Folder.java & Folder & 208 & 5 & ... & Structurer\\
			6 & k9mail-library.src.main.java.com.fsck.k9.mail.K9MailLib.java & K9MailLib & 67 & 7 & ... & Structurer\\
			7 & k9mail-library.src.main.java.com.fsck.k9.mail.Message.java & Message & 237 & 4 & ... & Coordinator\\
			8 &k9mail-library.src.main.java.com.fsck.k9.mail.Throttle.java & Throttle & 68 & 2 & ... & Controller \\
			9 & k9mail-library.src.main.java.com.fsck.k9.mail.K9.java & K9 & 43 & 0 & ... & Controller \\ 
			10 & k9mail-library.src.main.java.com.fsck.k9.mail.MailService.java & MailService & 25 & 2 & ... & Controller\\  
			\bottomrule
		\end{tabular*}
		\label{table:rs_sample} 
	\end{table*}
	
	\subsubsection*{Step 4: Data Integration}
	The methodology for creating a fine-grained dataset entails a systematic integration of preprocessed design smells and role-stereotype data, extending the approach originally proposed by \textcite{ogenrwot2021integration}. This integration technique focuses on combining information about design smells and role-stereotypes by aligning their common \textit{classpaths}. These matching classpaths act as essential identifiers, ensuring accurate association of relevant data from both datasets. To achieve this, we developed a simple algorithm to facilitate the integration process. Table \ref{table:ds_rs_processed} shows an example output of the fine-grained dataset generated by the algorithm. 
	\begin{algorithm}
		\caption{\enskip Integrate Design Semlls and Role Stereotype Data}\label{alg2}
		\begin{algorithmic}[1]
			\Procedure{IntegrateData}{$ds\_data, rs\_data$}       
			\State $result \leftarrow \{ \}$ 
			\State $ds\_classpaths \leftarrow ds\_data.get(classpath)$ 
			\State $rs\_classpaths \leftarrow rs\_data.get(classpath)$ 
			
			\For {$ (ds\_classpath$ \textbf{ or } $rs\_classpath ) $}
			\If{ $ds\_classpath = rs\_classpath$} 
			\State $result.add(classpath)$ 
			\State $result.add(ds\_data[design\_smells])$ 
			\State $result.add(rs\_data[label])$ 
			\EndIf
			\EndFor
			
			\State \Return $result$
			\EndProcedure
		\end{algorithmic}
		\label{al:data_integrated}
	\end{algorithm}
 The specifics of this algorithm are described below:
 \begin{enumerate}
\item {\it Initialization}: It begins by initializing an empty result set to hold the integrated information.
\item {\it Data Retrieval}: The algorithm retrieves the classpaths from both the design smells and role stereotypes data, storing these in separate lists.
\item {\it Iteration Over Classpaths}: It iterates over the union of classpaths from both datasets. During each iteration, it checks for matching classpaths between the design smells and role stereotypes.
\item \textit{Matching Process}: If a match is found, the algorithm adds the corresponding design smell data and role stereotype data to the result set alongside the matching classpath. This matching process is repeated until all classpaths in the combined set have been processed.
\item \textit{Result Compilation}: Upon completion of the iterations, the algorithm returns the integrated result set containing all matched entries.
 \end{enumerate} 
	\begin{table*}[h]
 \fontsize{9pt}{9pt}\selectfont
	\centering 
	\caption{Showing a sample of final dataset after integrating design smell and role stereotypes data.} 
	\begin{tabular*}{\textwidth}{@{\extracolsep\fill}llllllll@{\extracolsep\fill}}%
		
		\toprule
		\textbf{No.}& \textbf{FullClassPath} & \textbf{Classname}& \textbf{label} & \textbf{Blob} &\textbf{LongMethod} & \textbf{...}  \\ 
		\midrule 
		1 & k9mail-library.src.main.java.com.fsck.k9.mail.AuthType.java & AuthType& Service Provider & 3 & 1 & ... \\ 
		2 & k9mail-library.src.main.java.com.fsck.k9.mail.Address.java & Andress & Interfacer & 1 & 0  & ...  \\
		3 & k9mail-library.src.main.java.com.fsck.k9.mail.Body.java & Body & Interfacer & 0 & 0 &  ... \\
		4 &k9mail-library.src.main.java.com.fsck.k9.mail.Flag.java & Flag & Information Holder & 1 & 0 & ... \\ 
		5 & k9mail-library.src.main.java.com.fsck.k9.mail.Folder.java & Folder & Structurer & 1 & 0 & ... \\
		6 &k9mail-library.src.main.java.com.fsck.k9.mail.K9MailLib.java & K9MailLib & Structurer & 0 & 3 & ... \\
		7 & k9mail-library.src.main.java.com.fsck.k9.mail.Message.java & Message & Coordinator & 0 & 0 & ... \\
		8 &k9mail-library.src.main.java.com.fsck.k9.mail.Throttle.java & Throttle & Controller & 1 & 2 &  ...  \\
		9 & k9mail-library.src.main.java.com.fsck.k9.mail.K9.java & K9 & Controller & 4 & 0 & ... \\ 
		10 & k9mail-library.src.main.java.com.fsck.k9.mail.MailService.java & MailService & Controller & 1 & 2 & ... \\  
		\bottomrule
	\end{tabular*}
	\label{table:ds_rs_processed} 
\end{table*}
	
	\subsubsection*{Step 5: Data analysis}
 Using the fine-grained dataset obtained in {\it Step 4} (see Table \ref{table:ds_rs_processed}), we categorized each project by role-stereotype classifications to assess the prevalence of design smells. We calculated the percentage of classes with and without design smells and analyzed these patterns across different application domains (desktop vs. mobile). 
 Additionally, for each role-stereotype, the total number of design smells was computed to determine their percentage within each role stereotype. This was achieved by dividing the total number of design smells in each role-stereotype by the total number of design smells in the entire dataset. The selected projects were also grouped based on the application domain (desktop or mobile application). Then we conducted a Welch Two Sample t-test to assess the statistical significance of the observed differences in design smells between mobile and desktop applications. This test was specifically chosen because it accounts for the variations in sample sizes and variances between the two groups, ensuring a more accurate analysis given the noticeable disparities in the sizes of the selected mobile and desktop applications. This approach allows us to determine whether the differences in design smells are statistically significant, despite the inherent differences in application size and complexity. To gain a broader insight into the co-occurrence of design smells (which is interesting for {\it RQ1}), we conducted a Spearman rank correlation test and calculated the correlation coefficient ($R^2$) among the design smells. The forthcoming results section will offer detailed tables, figures, and statistical analyses that illustrate the interrelationships and distribution of design smells across various role stereotypes and application domains.

\subsubsection*{Step 6: Clustering and association rule mining}
In this study, we leveraged the Powered Outer Probabilistic Clustering (POPC) algorithm \cite{taraba2017clustering}. The selection of this algorithm was driven by the following considerations: First, numerous clustering algorithms, including the popular k-means algorithm, require the number of clusters to be specified in advance, which is a huge drawback. Some studies have employed methods such as the silhouette coefficient \cite{aranganayagi2007clustering,dinh2019estimating} and the elbow method \cite{bholowalia2014ebk, marutho2018determination}, among other approaches, to determine the optimal number of clusters. However, those methods have their limitations, for example: sometimes the elbow method fails to give a clear ``elbow point''. Second, k-means is not very suitable for a binary or sparse matrix. Our dataset is quite sparse, and since k-means depends on a distance measure (e.g euclidean), it becomes difficult to build cluster from a sparse matrix. 
 
 \noindent With POPC, there is no need to pre-define the number of clusters. The algorithm addresses this challenge through back-propagation i.e., start with many clusters and gradually optimize to obtain the optimal number of clusters. The algorithm is observed to work well on binary datasets and converges to the expected (optimal) number of clusters on theoretical examples as elaborated by \textcite{taraba2017clustering}. The algorithm can be summarized as follows:
	\begin{enumerate}
		\item Use k-means clustering to assign each sample $s_j$ to a cluster, denoted as $cl(s_j) = k$, where $k \in {1, ..., N}$ and $N$ is chosen to be half the number of data samples.
		\item Calculate $J_{r=0}$ to assess the initial clustering, with $r$ indicating the iteration of the algorithm.
		\item Increment $r$ to $r := r + 1$, initializing $J_r = J_{r-1}$.
		\item For each sample $s_j$, attempt to assign it to all clusters to which it does not currently belong, denoted as $(cl(s_j) \neq k)$.
		\begin{enumerate}
			\item [a)] If the temporary evaluation score $J_T > J_r$, then assign $J_r := J_T$ and move sample $s_j$ to the new cluster. 
		\end{enumerate}
		\item If $J_r$ is equal to $J_{r-1}$, then stop the algorithm. Otherwise, go back to step 3.
	\end{enumerate}

 \noindent We categorized the dataset into two groups: desktop and mobile projects, for the clustering task. This categorization was aimed at assessing potential differences in cluster formation between desktop and mobile applications. 
 To analyze role-stereotypes based on the design smells present within them, we restructured the clusters from the initial clustering output into a new formation. The process involved the following steps:
 \begin{enumerate}
 \item We utilized data from both desktop and mobile projects extracted from the initial clustering results to group clusters by role-stereotypes: For each cluster index (CID) within a role-stereotype, we assigned a value of $1$ to the corresponding cell in the role-stereotype's table, and $0$ otherwise. Design smells identified within that CID were also recorded. 
This action was repeated for all role-stereotypes.
\item We then extracted the columns of role-stereotypes to create a binary matrix. This matrix, formatted as an n-dimensional array, was inputted into a dendrogram creation tool. Using hierarchical clustering implemented with the Python Plotly package, we generated a dendrogram to illustrate the relationships between role-stereotypes based on design smells occurrence. The detailed findings are presented in the results section.
\end{enumerate}
	
\noindent In order to better understand the association of design smells with role-stereotypes, this study also explored an alternative approach to the clustering task. The study applied the well-known Apriori algorithms \cite{agrawal1996fast} to construct the association rules. Association rule discovery is an unsupervised learning technique used to detect local patterns which indicates attribute value conditions that occur together in a given dataset \cite{al2014improved}. The association rule mining task was carried out as follows;- We defined a set of items $I={i_1,...,i_n}$ which is a binary set of n attributes (design smells) and a set of m transaction $T={t_1,...,t_m}$, which indicate all Java classes analyzed. An association rule $X => Y$ where $X \subseteq I$ and $Y$ is a specific role-stereotype implies that a design smell (or group of design smells) $X$ is associated with $Y$ role-stereotype. We evaluate the strength of an association rule using three key metrics: support, confidence, and lift, detailed in their respective equations (\ref{eq:support}, \ref{eq:confidence} \& \ref{eq:lift}) shown below. Apriori algorithm was applied with the minimum support value empirically set to $0.05$.
	\\~\\
	\begin{equation}
		support (X => Y) = P (X, Y)\label{eq:support}
	\end{equation}
	\begin{equation}
		confidence = 	\frac{support (X \cup Y)}{support (X)} \label{eq:confidence}
	\end{equation}
	
	\begin{equation}
		lift (X => Y) = \frac{support (X \cup Y)}{support (X) \times support (Y)} \label{eq:lift}
	\end{equation}

\section{RESULTS}\label{sec:result}
	
	In this section, we present the results of our study. This is driven by answering the following research questions:
	\subsection*{RQ1: How do design smells vary across mobile and desktop applications? }
	
\begin{figure*}[!h]
	\centerline{\includegraphics[scale=0.7]{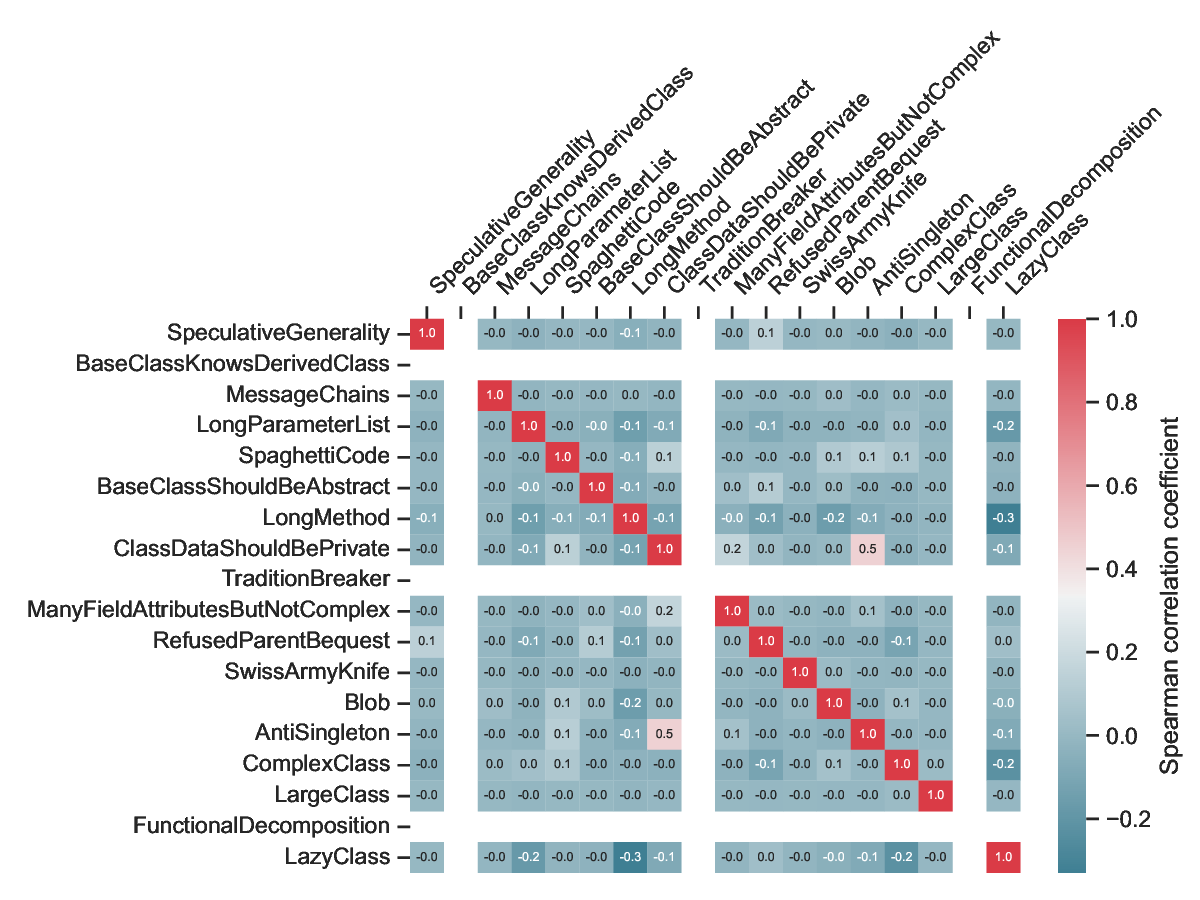}}
	\caption{Correlation analysis illustrating the co-occurrence patterns of design smells.\label{fig:headmapdsall}}
	\end{figure*}

\begin{figure*}[!h]
		\centerline{\includegraphics[scale=0.95]{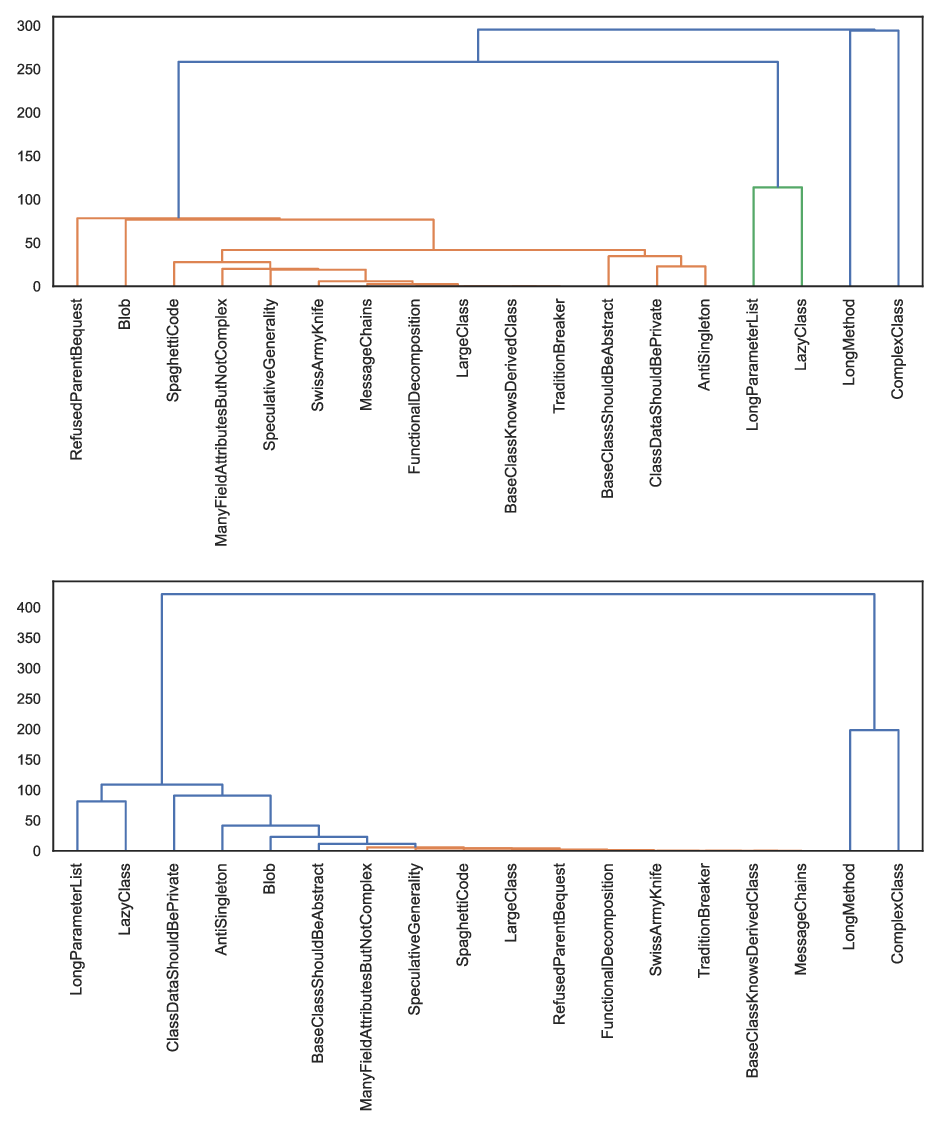}}
		\caption{Hierarchical cluster visualization of design Smells that frequently co-occur in desktop (top plot) and mobile (bottom plot) applications.\label{fig:dendogram_ds}}
	\end{figure*}

	In our prior study \cite{ogenrwot}, we focused on contrasting the prevalence of design smells between desktop and mobile applications. Specifically, we explored how the application type (desktop or mobile) influences the diversity, distribution, and magnitude of design smell occurrences. However, the study was constrained by a relatively small dataset, and the hypothesis testing was conducted considering the entire dataset without a more granular approach. In this study, we build on our prior work by investigating the density of design smells in mobile and desktop applications. This involves comparing the number of design smells per thousand lines of code (KLOC) in the selected projects. Notably, the average number of design smells per KLOC in desktop applications is relatively higher (37.3\%) compared to the mobile applications (32.7\%). Furthermore, we conducted Welch Two Sample t-test, to assess the statistical significance of these observed differences. The choice of the Welch Two Sample t-test is motivated by the noticeable variations in the sizes of the selected mobile and desktop applications. The result of the Welch's Two Sample t-test indicates that the aforementioned difference is not statistically significant  with the $t$ and $p$ values of $0.535$ and $0.597$ respectively. Recall that in Section \ref{sec:method} ({\it Step 4}) we calculated the correlation coefficient ($R^2$) among design smells, as depicted in Figure \ref{fig:headmapdsall}. The results indicate a generally low correlation between design smells, with the highest observed value being 0.5, which pertains to the relationship between \textit{AntiSingleton} and \textit{ClassDataShouldBePrivate}.
	Using the POPC clustering algorithm, this study affirms theories pertaining to shared characteristics and similarities among design smells. The hierarchical clustering, as illustrated by the dendrogram shown in Figure \ref{fig:dendogram_ds}, unveils the co-occurrence of design smells in both desktop and mobile applications. The results from Figure \ref{fig:dendogram_ds} are further elaborated in Table \ref{table:ds_den_table}, presenting an insight into the co-occurrence patterns of design smells in desktop and mobile applications.
	
	To compliment POPC clustering algorithm, the study explored an alternative approach based on association rule mining. Using the apriori algorithm, association rules were established to uncover relationships between design smells and their corresponding degrees of confidence, as detailed in Table \ref{table:assoc_ds_ds}. The hyper-parameter of the apriori algorithm was fine-tuned by setting the minimum support to 0.05. The results from the association rule mining, presented in Table \ref{table:assoc_ds_ds}, showcases 12 association rules ordered in descending order of confidence. Notably, the study observed a high-confidence association between \textit{AntiSingleton} and \textit{ClassDataShouldBePrivate} with a confidence level of 0.66. Conversely, the association of \textit{LongMethod} with \textit{LongParameterList} and \textit{ComplexClass} exhibited the lowest confidence, registering at 0.10.
		\begin{table*}[h!]
		\centering 
		\caption{The  association of design smells with each other and their respective degrees of confidence.} 
		\begin{tabular*}{453pt}{@{\extracolsep\fill}llr@{\extracolsep\fill}}%
			
			\toprule
			\textbf{No} & \textbf{Rule} &\textbf{ Confidence}  \\
			\midrule
			1. &(AntiSingleton) $\rightarrow$ (ClassDataShouldBePrivate) & 0.66 \\
			2. &(LongMethod, LongParameterList)	 $\rightarrow$ (ComplexClass) & 0.57 \\
			3. &(LongParameterList, ComplexClass) $\rightarrow$ (LongMethod)  & 0.50 \\
			4. &(ComplexClass) $\rightarrow$ (LongMethod) & 0.46 \\
			5. &(LongParameterList)	 $\rightarrow$ (ComplexClass) & 0.41 \\
			6. &(LongMethod) $\rightarrow$ (ComplexClass) & 0.37 \\
			7. &(ClassDataShouldBePrivate) $\rightarrow$ (AntiSingleton) & 0.49 \\
			8. &(LongMethod, ComplexClass) $\rightarrow$ (LongParameterList) & 0.25 \\
			9. &(ComplexClass) $\rightarrow$ (LongParameterList) & 0.24 \\
			10. &(LongParameterList)	 $\rightarrow$ (LongMethod, ComplexClass) & 0.20 \\
			11. &(ComplexClass)	 $\rightarrow$ (LongMethod, LongParameterList) & 0.12 \\
			12. &(LongMethod) $\rightarrow$ (LongParameterList, ComplexClass) & 0.10 \\
			\bottomrule
		\end{tabular*}
		\label{table:assoc_ds_ds}
	\end{table*} 
	
\begin{table}[h!]
\centering
    \caption{The co-occurrence of design smells across desktop and mobile applications extracted from Figure \ref{fig:dendogram_ds}.} 
    \label{table:ds_den_table}
    \begin{tabularx}{453pt}{lX} 
        \toprule
        \textbf{No} & \textbf{Co-occurrence Set in Desktop} \\
        \midrule
        1. & \{BaseClassShouldBeAbstract, ManyFieldAttributeButNotComplex\} \\ 
        2. & \{Speculative Generality, FunctionalDecomposition, LargeClass, BaseClassKnowsDerivedClass, TraditionalBreaker\}  \\ 
        3. & \{SwissArmyKnife, MessageChains, SpaghettiCode\} \\ 
        4. & \{Blob, ComplexClass\} \\ 
        5. & \{ClassDataShouldBePrivate, AntiSingleton\} \\ 
        6. & \{LongParameterList, LongMethod\} \\
        \bottomrule
    \end{tabularx}

    \vspace{5mm} 

    \begin{tabularx}{453pt}{lX} 
        \toprule
        \textbf{No} & \textbf{Co-occurrence Set in Mobile}  \\
        \midrule
        1. & \{LongMethod, LazyClass\} \\
        2. & \{BaseClassShouldBeAbstract, Blob\} \\ 
        3. &  \{FunctionalDecomposition, SwissArmyKnife, RefusedParentBequest, SpaghettiCode, BaseClassKnowsDerivedClass, MessageChains\} \\ 
        4. & \{ManyFieldAttributesButNotComplex, LargeClass\} \\ 
        5. & \{LongParameterList, AntiSingleton\} \\ 
        6. &  \{ClassDataShouldBePrivate, ComplexClass\} \\ 
        \bottomrule
    \end{tabularx}
\end{table}

\begin{tcolorbox}
\faLightbulbO \hspace{2mm}\textbf{Summary of Results for RQ1:} 
\begin{enumerate}[leftmargin=*]
    \item On average, design smells are relatively higher in desktop applications than in mobile applications. 
    \item The average difference in the number of design smells in desktop and mobile applications is not statistically significant.
    \item The hierarchical clustering and association rule mining reveals co-occurrence among design smells.
    \end{enumerate}
\end{tcolorbox}

\subsection*{RQ2: How do design smells relate with different types of role-stereotypes?}
\begin{figure*}[!h]
		\centering
		\includegraphics[scale=0.6]{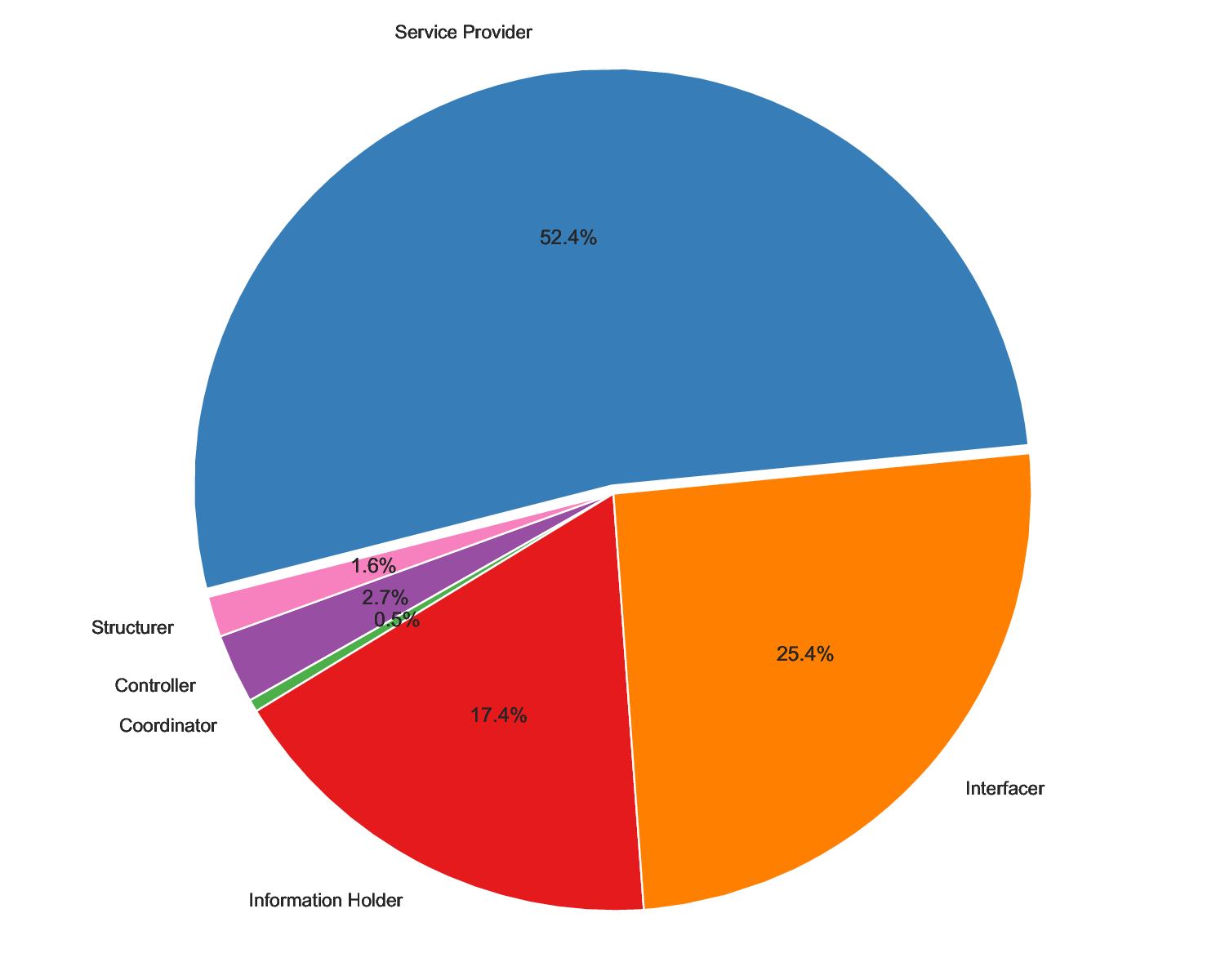}
		\caption{Distribution of design smells in each role-stereotype.}
		\label{fig:pie_chart_percentages}
	\end{figure*}

	In this second research question, the main objective was to understand the relationship between responsibility (role) stereotypes and design smells. To address this research question, Tables \ref{table:tb_percentages-desktop} and \ref{table:tb_percentages-mobile} provide insights into the percentages of classes containing design smells and those without, categorized by specific role stereotypes. Notably, the {\tt Service Provider} responsibility stereotype consistently exhibits the highest percentage of classes with design smells in both desktop and mobile applications. This finding is also supported by our aggregated results in Figure \ref{fig:pie_chart_percentages}, where we present the overall percentages of design smells within each role stereotype. Specifically, design smells tend to be more prevalent in {\tt Service Provider} (52.4\%), {\tt Interfacer} (25.4\%), and {\tt Information Holder} (17.4\%) compared to {\tt Controller} (2.7\%), {\tt Structurer} (1.6\%), and {\tt Coordinator} (0.5\%) role stereotypes.
	
	\begin{figure*}[!h]
		\centerline{\includegraphics[scale=0.7]{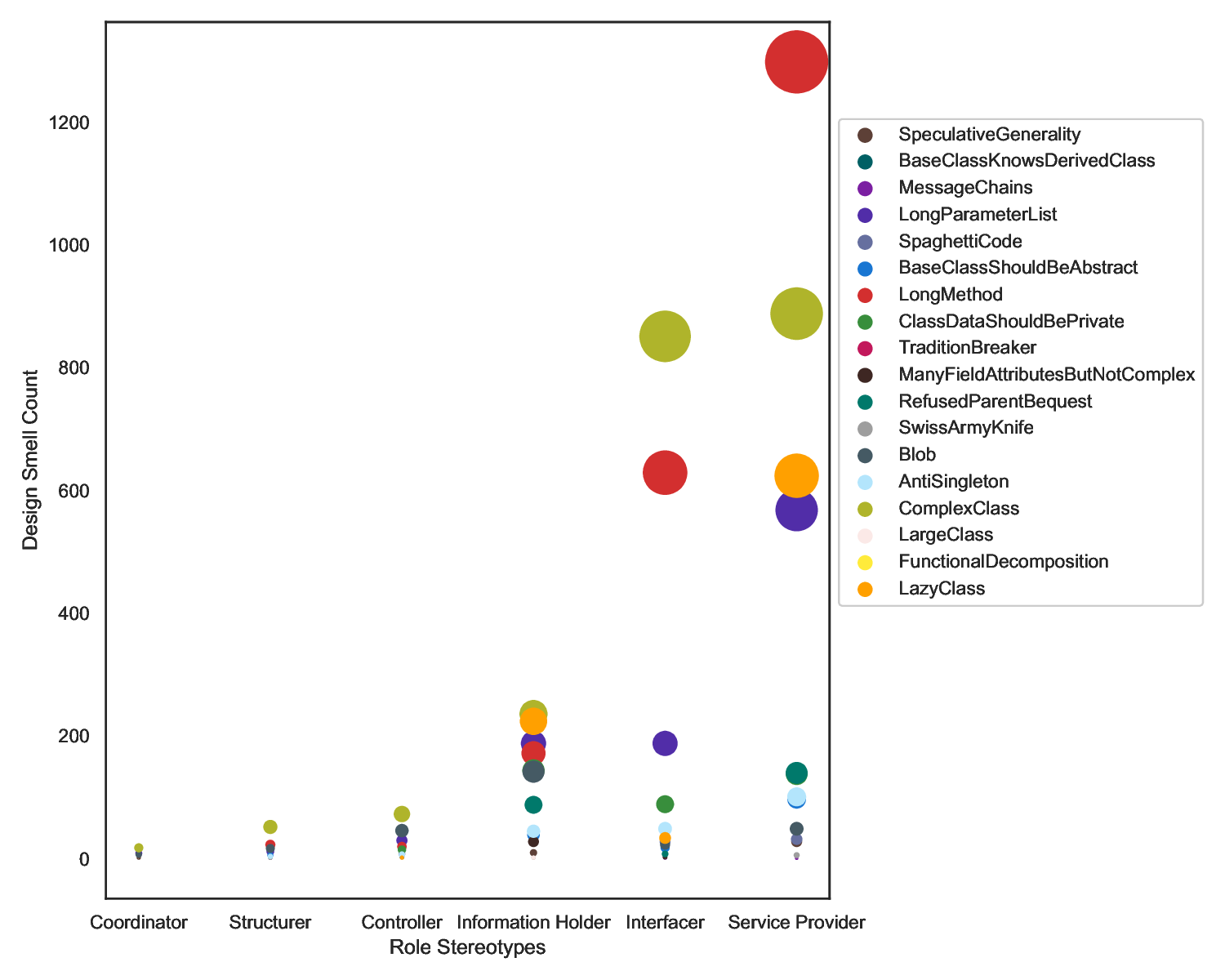}}
		\caption{Bubble chart showing the distribution of design smells in each category role-stereotype. The bubble size corresponds to the frequency of occurrence of a specific design smell.\label{fig:bubble_chart}}
	\end{figure*}

\begin{table*}[h]%
\fontsize{8pt}{8pt}\selectfont
	\caption{Provides a summary of the percentages of classes in {\bf desktop} applications categorized by role-stereotypes. It distinguishes between classes with design smells and those without design smells in relation to the total classes within the project. The highlighted cells draw attention to role-stereotypes exhibiting the highest percentage of classes with design smells for a specific project.\label{table:tb_percentages-desktop}}
	\begin{tabular*}{\textwidth}{@{\extracolsep\fill}llllllllllllll@{}}
		\toprule
		&&\multicolumn{2}{@{}l}{\textbf{SP}} & \multicolumn{2}{@{}l}{\textbf{CO}} & \multicolumn{2}{@{}l}{\textbf{IH}}& \multicolumn{2}{@{}l}{\textbf{IT}}& \multicolumn{2}{@{}l}{\textbf{CT}}& \multicolumn{2}{@{}l}{\textbf{ST}} \\\cmidrule{3-4}\cmidrule{5-6}\cmidrule{7-8}\cmidrule{9-10}\cmidrule{11-12}\cmidrule{13-14}
		\textbf{Projects} & \textbf{NOC} & \textbf{a (\%)}  & \textbf{b (\%)}  & {\textbf{a (\%)}}  & \textbf{b (\%)}  & \textbf{a (\%)}  & \textbf{b (\%)} & \textbf{a (\%)}  & \textbf{b (\%)} & \textbf{a (\%)}  & \textbf{b (\%)} & \textbf{a (\%)}  & \textbf{b (\%)}  \\
		\midrule
		SweetHome3D& 546  & 2.93       & 26.19      &\textbf{5.31} & 1.65 & 3.11                         & 38.46 & 2.56                         & 8.97  & 0.73       & 3.11       & 1.65       & 5.31   \\
		Mars Simulation & 1109  &\textbf{19.93 }      & 23.81      & 0.45 & - & 7.75                         & 23.81 & 12.35                         & 9.47  & -       & 0.09       & 0.90       & 1.44       \\ 
		ArogUML & 1236 &\textbf{28.07}       & 46.36  & 0.16 & - & 6.07                         & 10.76 & 6.23                         & 1.70  & 0.08       & 0.16       & 0.24       & 0.16       \\ 
		Edit & 577  & \textbf{46.45}       & 24.78      & 0.52 & - & 13.69                         & 0.32 & 5.03                         & 0.87  & -      & -       & 0.17       & 0.17       \\ 
		Gantt Project & 671     &\textbf{21.31 }      & 46.05      & 0.15 & - & 6.71                        & 14.90  & 7.60                         & 1.04  & -       & 0.45       & 1.64	       & 0.15       \\ 
		GoGreen & 60  & \textbf{31.67}      & 35.00     & -	 & 1.67 & 15   &  15            & -  & -       & -       & 1.67	       & -   & -   \\
		LiveChat Server & 23 &\textbf{26.09}     & 21.74     & -	 & - & 21.74  & 13.04  & 17.39	& -  & -       & -       & -	       & -      \\
		Checkstyle & 1008  & \textbf{9.62}  & 46.33   & -	 & - & 7.14  & 34.42 & -            & 1.79  & -       & 0.10       & -	       & 0.60 \\
		Keystore explorer & 400 & \textbf{25.75}  & 29.75   & -	 & - & 5.50 & 20.00 & 14.00  & 4.25  & -  & 0.75 & - & -      \\
		Angry IP Scanner & 159 &\textbf{25.16}	& 49.06  & -	 & - & 5.03  & 11.95  & 7.55      & 0.63  & -       & -       & 0.63 & -      \\ 
		jetUML & 173 & \textbf{38.73} & 24.86 & 0.58 & - & 9.83                       & 20.23 & 4.62 & -  & -       & 1.16 & -	       & -      \\
		jPass & 38  &  \textbf{13.16} & 55.26 & - & - & 5.26  & 18.42 & 5.26 & 2.63 & -       & -       & -	       & -      \\
		LogFX & 44  & \textbf{25} & 27.7 & - & - & 6.82 & 25.00 & 9.09 & 6.82 & -       & -       & -	       & -      \\
		PGP Tool & 226  & \textbf{12.83} & 37.17 & - & - & 6.19  & 26.55 & 10.18 & 7.08 & -       & -       & -	       & -      \\
		Freemind & 370  &  \textbf{22.70} & 45.95 & 1.89 & 0.54 & 7.30 & 10.27 & 6.49 & 2.97 & -       & -       & 1.89	       & -      \\
		\bottomrule
	\end{tabular*}
		\begin{tablenotes}
			\item[a: ] a: Percentage of classes containing design smells
			\item[b: ] b: Percentage of classes that does not contain design smells
		\end{tablenotes}
\end{table*}

\begin{table*}[h]%
\fontsize{8pt}{8pt}\selectfont
	\caption{Provides a summary of the percentages of classes in {\bf mobile} applications categorized by role-stereotypes. It distinguishes between classes with design smells and those without design smells in relation to the total classes within the project. The highlighted cells draw attention to role-stereotypes exhibiting the highest percentage of classes with design smells for a specific project.\label{table:tb_percentages-mobile}}
	\begin{tabular*}{\textwidth}{@{\extracolsep\fill}llllllllllllll@{}}
		\toprule
		&&\multicolumn{2}{@{}l}{\textbf{SP}} & \multicolumn{2}{@{}l}{\textbf{CO}} & \multicolumn{2}{@{}l}{\textbf{IH}}& \multicolumn{2}{@{}l}{\textbf{IT}}& \multicolumn{2}{@{}l}{\textbf{CT}}& \multicolumn{2}{@{}l}{\textbf{ST}} \\\cmidrule{3-4}\cmidrule{5-6}\cmidrule{7-8}\cmidrule{9-10}\cmidrule{11-12}\cmidrule{13-14}
		\textbf{Projects} & \textbf{NOC} & \textbf{a (\%)}  & \textbf{b (\%)}  & {\textbf{a (\%)}}  & \textbf{b (\%)}  & \textbf{a (\%)}  & \textbf{b (\%)} & \textbf{a (\%)}  & \textbf{b (\%)} & \textbf{a (\%)}  & \textbf{b (\%)} & \textbf{a (\%)}  & \textbf{b (\%)}  \\
		\midrule
		\begin{tabular}[c]{@{}l@{}}K9 \\ Mail\end{tabular}  & 779   & 4.36       & 37.10      & 0.90                         & 1.67 & \textbf{4.62} & 25.03 & 2.57                         & 7.32  & 2.05       & 8.09       & 1.41       & 4.88       \\
		Bitcoin Wallet  & 222 & 2.70       & 22.97      & 0.90                         & 1.35 & 4.05                         & 33.33 & \textbf{6.31} & 21.62 & -        & 0.90       & 1.8        & 4.05       \\
		Keepassdroid & 211 & \textbf{29.86}       & 39.34      & - & - & 9.95	                        & 14.22  & 6.16                         & 0.47  & -       & -       & -	       & -     \\ 
		OpentripPlanner  & 53 &\textbf{24.53}       & 43.40      & - & - & 3.77                       & 18.87   & 5.66	                         & 3.77  & -       & -       & -	       & -       \\ 
		Tweet Lanes & 130   &\textbf{22.31}       & 33.08      & - & - & 8.46	                       & 15.38   & 15.38                        & 4.62   & -      & -       & -	       & -       \\
		Text Secure & 1332 & \textbf{23.87}       & 42.64      & 0.15	 & - & 3.83                        & 11.41  & 12.24            & 4.80  & 0.08       & 0.68       & 0.30	       & -       \\
		Telegram & 679  & 19.15       & 20.91      & 2.21	 & 0.29 & 6.33                        & 9.13  &\textbf{25.04}            & 16.05  & -       & 0.15       & 0.44	       & 0.29 \\
		Materialistic& 131 & \textbf{25.19} & 42.75 & -	 & - & - & 15.27 & 12.21 & 3.05 & 1.53  & - & - & - \\
		Telecine & 23  &  \textbf{26.09} & 26.09 & - & - & -                        & 30.43  & 13.04  & 4.35  & - & - & -	& - \\
		Amaze File Manager & 265   & \textbf{23.02} & 38.87 & -	 & - & 6.79	& 11.32 & 16.60 & 3.40 & -       & -       & -	       & -      \\
		Omni-Notes & 159  & \textbf{22.64}	& 45.91  & -	 & 0.63 & 8.81 & 10.69 & 8.81 & 2.52 & -       & -       & -	       & -      \\
		AntennaPod  & 387  & \textbf{18.60}  & 43.93 & 0.26 & - & 3.62 & 16.02  & 10.85 & 6.20 & -  & 0.52 & -	 & - \\ 
		GnuCash & 147  &  \textbf{30.61} & 35.37   & - & - & 5.44 & 8.16 & 15.65 & 4.76 & -  & - & -  & -  \\
		Timber  & 123  & \textbf{24.54} & 36.20 & -	 & - & 11.66 & 9.82 & -          11.66  & 4.29  & -       & 1.23      & -	       & 0.61      \\
		SeeWeather & 29  & \textbf{31.03}	& 31.03  & -	 & - & 24.14 & 3.45 & 10.34 & -  & -       & -       & -	       & -      \\
		\bottomrule
	\end{tabular*}
	\begin{tablenotes}
		\item[a: ] a: Percentage of classes containing design smells
		\item[b: ] b: Percentage of classes that does not contain design smells
	\end{tablenotes}
\end{table*}

	The second part of this research question explores the prevalence of specific design smells within each role-stereotype. Figure \ref{fig:bubble_chart} illustrates (i) the occurrence of particular design smells in the respective role-stereotypes and (ii) the magnitude of their occurrence, depicted by the height of the stacked bar plot. Consistent with earlier observations, design smells are notably frequent in the {\tt Service Provider}, {\tt Interfacer}, and {\tt Information Holder} role-stereotypes. Furthermore, {\tt LongMethod}, {\tt ComplexClass}, and {\tt LongParameterList} exhibit the highest occurrence frequency across all role-stereotype categories, surpassing {\tt LargeClass}, {\tt MessageChains}, and {\tt SpaghettiCode}. Additionally, it was noted that {\tt Service Provider} and {\tt Information Holder} are susceptible to a broader spectrum of design smells compared to other role-stereotypes, such as {\tt Coordinator}, as depicted in Table \ref{table:tb_highlighted}.
		\begin{table*}[h]
		\centering 
		\caption{The association of various design smells with role-stereotypes with the respective degrees of confidence.} 
		\begin{tabular*}{453pt}{@{\extracolsep\fill}llr@{\extracolsep\fill}}%
			
			\toprule
			\textbf{No} & \textbf{Rule} &\textbf{ Confidence}  \\
			\midrule
			1. & (ComplexClass) $\rightarrow$ (Service Provider) & 0.72 \\
			2. & (ComplexClass) $\rightarrow$ (Interfacer) & 0.63 \\
			3. & (LongMethod) $\rightarrow$ (Interfacer) & 0.62 \\
			4. & (LongMethod) $\rightarrow$ (Service Provider) & 0.60 \\
			5. & (LongMethod, LongParameterList) $\rightarrow$ (Service Provider) & 0.58 \\
			6. & (LongParameterList) $\rightarrow$ (Service Provider) & 0.57 \\
			7. & (LongMethod, ComplexClass) $\rightarrow$ (Service Provider) & 0.49 \\
			8. & (LongParameterList, ComplexClass) $\rightarrow$ (Service Provider) & 0.46 \\
			9. & (LazyClass) $\rightarrow$ (Service Provider) & 0.45 \\
			10. & (LongMethod, ComplexClass) $\rightarrow$ (Interfacer) & 0.40 \\
			11. & (ComplexClass) $\rightarrow$ (Interfacer) & 0.38 \\
			12. & (ClassDataShouldBePrivate) $\rightarrow$ (Information Holder) & 0.36 \\
			13. & (LongParameterList, ComplexClass) $\rightarrow$ (Interfacer) & 0.36 \\
			14. & (LongParameterList) $\rightarrow$ (Information Holder) & 0.26 \\ 
			\bottomrule
		\end{tabular*}
		\label{table:assoc_ds_rs}
	\end{table*} 
	
	\textbf{Association Rule Mining:} The results of association rule mining are presented in Table \ref{table:assoc_ds_rs}. Evidently, associations between design smells and specific role stereotypes are observable. These associations can be reasonably attributed to the inherent characteristics and collaborative nature of role stereotypes. For example, we have noticed a connection between the Service Provider role-stereotype and the ComplexClass design smell. Although a Service Provider class might not inherently display the ComplexClass design smell, it could potentially contribute to it depending on its implementation and usage within an application. This could occur if: (1) the service provider class grows too large and begins to handle multiple responsibilities or functionalities unrelated to the service it offers, or (2) the service provider class becomes tightly coupled with other classes or components in the system. Additionally, the study uncovers that design smells such as LongMethod, LongParameterList, and ComplexClass are commonly associated with multiple role stereotypes.
		\begin{tcolorbox}
		\faLightbulbO \hspace{2mm}\textbf{Summary of Results for RQ2:}
			\begin{enumerate}[leftmargin=*]
				\item Design smells tend to occur more often in {\tt Service Provider} (53.4\%), {\tt Interfacer} (26.7\%) and \textit{Information Holder} (15.1\%) than in \textit{Controller} (2.6\%), \textit{Structurer} (1.7\%) and \textit{Coordinator} (0.5\%) role-stereotypes.
				\item The study observed that \textit{LongMethod}, \textit{ComplexClass} and \textit{LongParameterList} have the highest frequency of occurrence across the entire role-stereotype categories compared to \textit{LargeClass}, \textit{MessageChains} and \textit{SpaghettiCode}.
				\item There exist, association between design smells and some role-stereotypes. It is believed that these associations are attributed to the characteristics and collaborative nature of role-stereotypes.
			\end{enumerate}

    \end{tcolorbox}
	
	\subsection*{RQ3: Does the type of application (desktop or mobile) influence the relation between design smells and role-stereotypes?}
	
\begin{figure*}[h]
		\centering
		\includegraphics[scale=0.6]{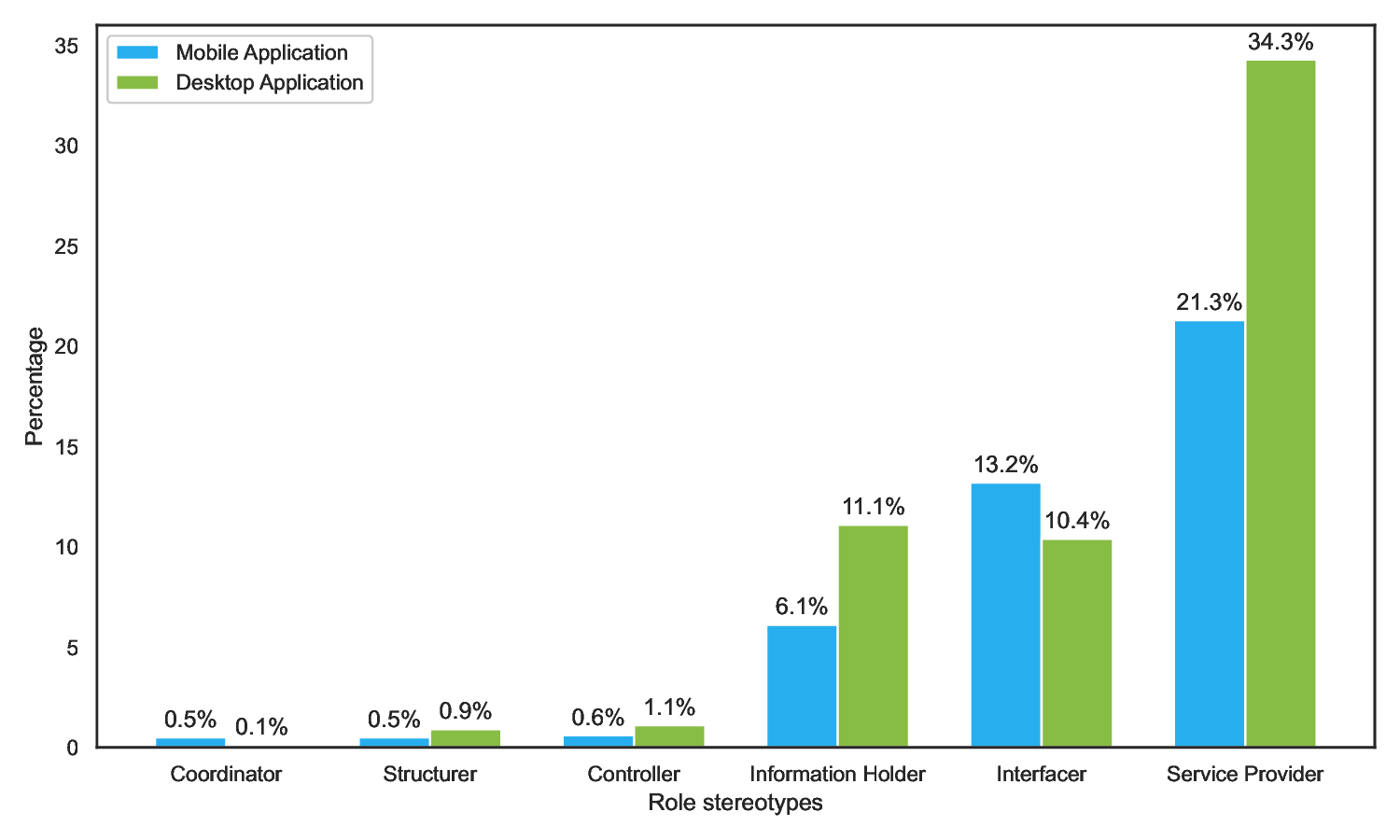}
		\caption{Percentage of design smells in mobile vs desktop application for each role-stereotype.}
		\label{fig:bar_chart_percentage}
	\end{figure*}

	To address this question, we start with a comparison of the percentage of design smells in mobile and desktop applications within each role stereotype, as shown in Figure \ref{fig:bar_chart_percentage}. This visualization not only underscores the higher prevalence of design smells in desktop applications but also identifies role stereotypes of particular interest. We observed design smells are more prevalent in {\tt Service Provider}, {\tt Information Holder}, {\tt Controller}, and {\tt Structurer} role stereotypes across both desktop and mobile applications. Conversely, {\tt Interfacer} and {\tt Coordinator} role stereotypes exhibit a higher occurrence of design smells in mobile than desktop applications. The detailed breakdown in Table \ref{table:tb_highlighted} provides insights into the frequency and variety of design smells, highlighting that {\tt Service Provider} and {\tt Information Holder} role stereotypes exhibit the highest diversity of design smells. Additionally, specific design smells are observed to occur exclusively in particular role stereotypes and application types.
	
		\begin{table*}[h]%
  \fontsize{8pt}{8pt}\selectfont
			\caption{Distribution of design smells in each role-stereotype across mobile and desktop applications. The cells with tick mark indicate the presence of a particular design smell in a given role-stereotype and type of application. The role-stereotypes are abbreviated as follows: Coordinator (CO), Structurer (ST), Service Provider (SP), Information Holder (IH), Controller (CT) and Interfacer (IT).\label{table:tb_highlighted}}
			\begin{tabular*}{\textwidth}{@{\extracolsep\fill}lllllllllllll@{}}
				\toprule
				&\multicolumn{2}{@{}l}{\textbf{CO}} & \multicolumn{2}{@{}l}{\textbf{ST}} & \multicolumn{2}{@{}l}{\textbf{CT}}& \multicolumn{2}{@{}l}{\textbf{IH}}& \multicolumn{2}{@{}l}{\textbf{IT}}& \multicolumn{2}{@{}l}{\textbf{SP}} \\\cmidrule{2-3}\cmidrule{4-5}\cmidrule{6-7}\cmidrule{8-9}\cmidrule{10-11}\cmidrule{12-13}
				\textbf{Design Smells} & \textbf{M}  & \textbf{D}  & {\textbf{M}}  & \textbf{D}  & \textbf{M}  & \textbf{D} & \textbf{M}  & \textbf{D} & \textbf{M}  & \textbf{D} & \textbf{M}  & \textbf{D}  \\
				\midrule
				AntiSingleton  & - & - & \checkmark& \checkmark & \checkmark & \checkmark & \checkmark & \checkmark  &  \checkmark & - & \checkmark & \checkmark \\ 
				
				BaseClassKnowsDerivedClass & -  & - & - & - & - & - & - & - & - & - & - & - \\ 
				
				BaseClass ShouldBePrivate & - & - & \checkmark & - & - & \checkmark & \checkmark & \checkmark & \checkmark & \checkmark & \checkmark & \checkmark \\ 
				
				Blob  & \checkmark & \checkmark & \checkmark & \checkmark& \checkmark & \checkmark& \checkmark & \checkmark& \checkmark & \checkmark& \checkmark & \checkmark \\ 
				
				ClassDataShouldBePrivate& - & - & \checkmark & \checkmark & \checkmark & \checkmark & \checkmark & \checkmark & \checkmark & \checkmark &  \checkmark & \checkmark  \\ 
				
				ComplexClass &  \checkmark & \checkmark & \checkmark & \checkmark& \checkmark & \checkmark& \checkmark & \checkmark& \checkmark & \checkmark& \checkmark & \checkmark \\ 
				
				FunctionalDecomposition  & - & - & - & - & - & - & - & - & - & - & - & - \\ 
				
				LargeClass  & - & - & - & - & - & - & \checkmark & - & - & - & - & - \\ 
				
				LazyClass & - & - & - & - & - & \checkmark & \checkmark & \checkmark & \checkmark & \checkmark & \checkmark & \checkmark \\ 
				
				LongMethod & - & - & \checkmark & \checkmark & \checkmark &  \checkmark & \checkmark & \checkmark &  \checkmark & \checkmark &  \checkmark & \checkmark \\ 
				
				LongParameterList & \checkmark & \checkmark & \checkmark & \checkmark& \checkmark & \checkmark& \checkmark & \checkmark& \checkmark & \checkmark& \checkmark & \checkmark \\ 
				
				MessageChains & -  & - & - & \checkmark & - & - & - & - & - & \checkmark & - & \checkmark \\ 
				
				ManyFieldAttributesButNotComplex & - & - & - & - & - & - & \checkmark & \checkmark & - & - & - & - \\ 
				
				RefusedParentBequest & - & - & - & - & - & - & - & \checkmark & - & \checkmark & - & \checkmark \\
				
				SpaghettiCode  & - & - & - & - & - & \checkmark & - & - & \checkmark & \checkmark  & - & \checkmark  \\ 
				
				SpeculativeGenerality &  \checkmark & - & - & -  & - & - & \checkmark & \checkmark & - & - & \checkmark & \checkmark \\ 
				
				SwissArmyKnife  & - & - & -  & \checkmark & - & - & - & \checkmark& - & - & - & \checkmark \\
				
				TraditionBreaker & - & - & - & - & - & - & - & - & - & - & - & - \\ 
				\bottomrule
			\end{tabular*}
			\begin{tablenotes}
				\item[M: ] M: Mobile application.
				\item[D: ] D: Desktop application.
			\end{tablenotes}
		\end{table*}
		
\begin{figure*}[h]
		\centering
		\includegraphics[scale=0.6]{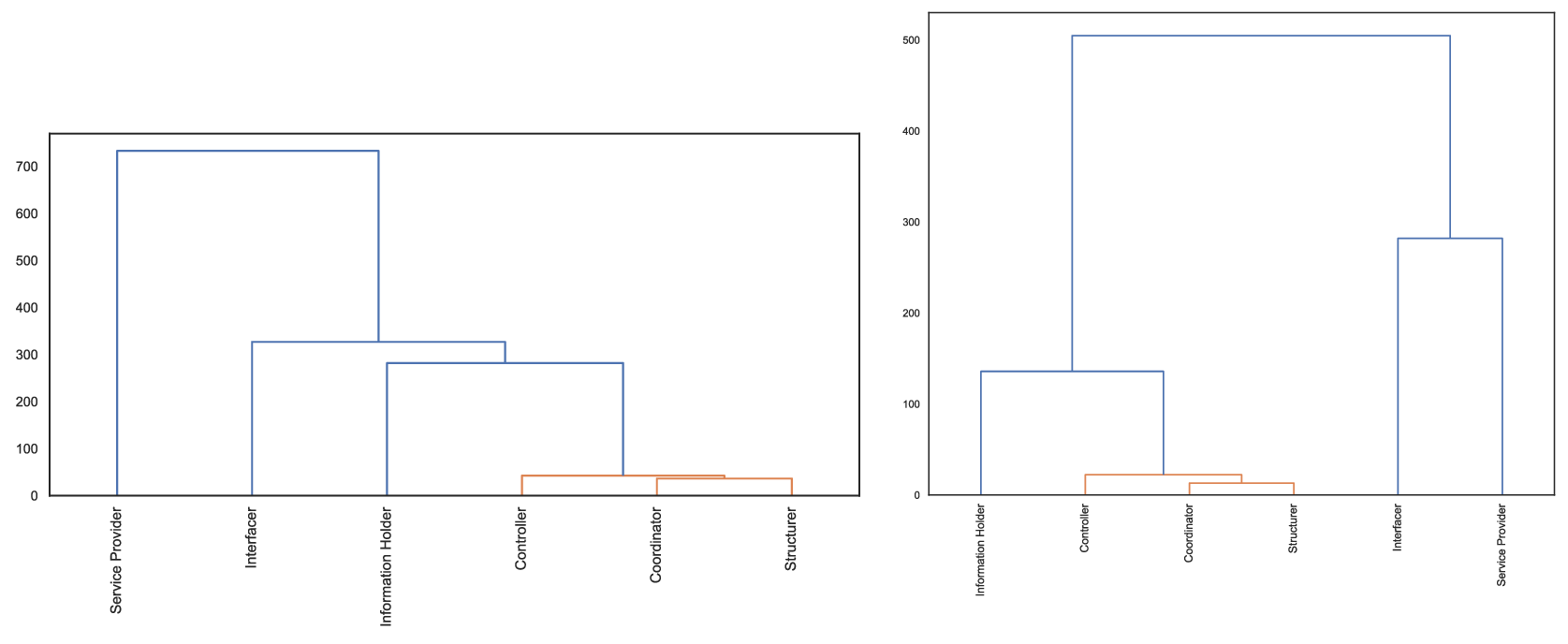}
		\caption{Hierarchical clusters represented by a dendrogram showing groups of role-stereotype in desktop (left) and mobile (right) applications with common type of design smells.}
		\label{fig:dendogram_rs}
	\end{figure*}

	Finally, we studied the distribution of design smells using unsupervised learning. The objective was to observe the groups/pairs of role-stereotypes which exhibit similar type of design smells and to study whether this relation stretches across mobile and desktop application. Figure \ref{fig:dendogram_rs} indicates that the groups \{{\tt Coodinator, Structurer}\}, 
	\{{\tt Controller, nterfacer}\} and 
	\{{\tt Servicer Provider, Information Holder}\} are quite similar in term of the design smells that occur within the scope of desktop application. The same figure also shows clusters of role-stereotypes:  \{{\tt Interfacer, Service Provider, Information Holder}\} and \{{\tt Coordinator, Controller, Structurer}\} across mobile applications. 
	We noticed one difference in these groupings across desktop and mobile applications. Specifically, the {\tt Interfacer} role-stereotype has sightly different types of design smells in desktop applications as compared to mobile applications. The fact that interfacers might be designed somewhat differently in the mobile application also surfaced earlier where a hypothetical explanation is the different availability of libraries and frameworks in the mobile application.
	A similarity across mobile and desktop is that both {\tt Service Provider} and {\tt Information Holder} are in both cases grouped as 'similar' in terms of occurrences of design smells.
	
	\begin{figure*}[h]
		\centering
		\includegraphics[width=\textwidth]{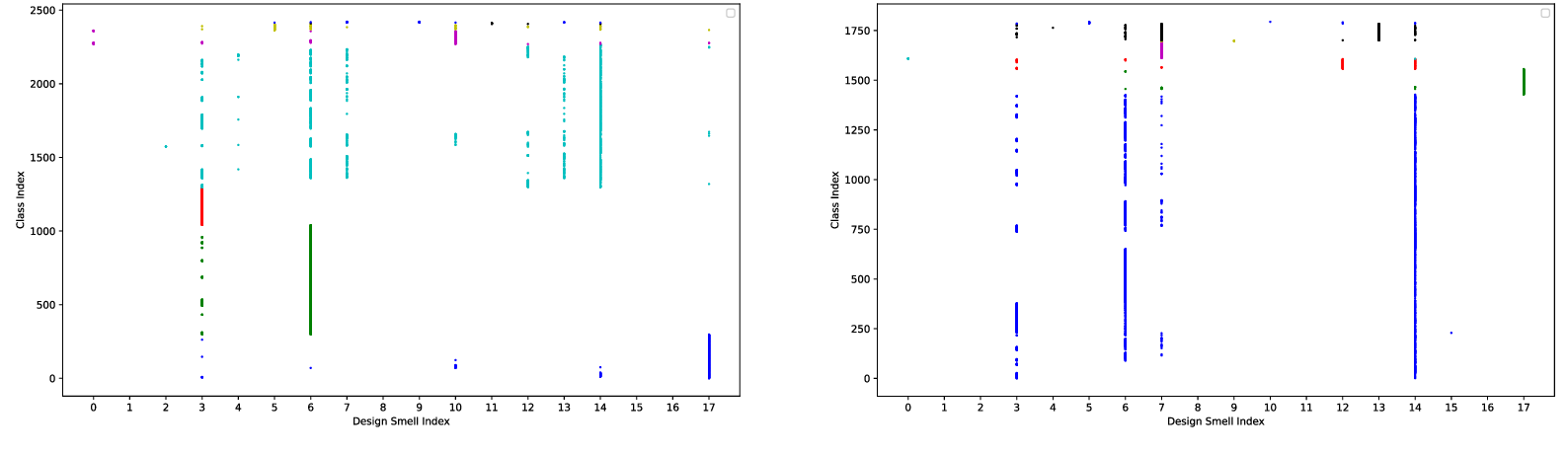}
		\caption{Results of clustering created with the POPC algorithm from desktop (left) and mobile (right) projects respectively. The vertical axis represents different samples (classes) belonging to different clusters (indicated in different colours) and the horizontal axis shows different features (design smells).}
			\label{fig:popc_all}
	\end{figure*}

	\textbf{Association Rule Mining:}
	The  results of association rule mining are presented in Table \ref{table:assoc_rs}. This finding is relatively consistent with the results of POPC algorithms reflected in Figure \ref{fig:dendogram_rs}. Based on association rule mining, we can observe that there are generally strong associations of the different role stereotypes with each other across desktop and mobile applications. For example, the study noticed the following association rule with 100\% confidence in desktop applications; \{IT $\rightarrow$ SP\}, \{ST $\rightarrow$ SP\} and \{IT, CT $\rightarrow$ SP\}. Other generated rules such as \{CT $\rightarrow$ SP\}, \{IT $\rightarrow$ CT\}, \{IT, SP $\rightarrow$ CT\}, \{SP, CT $\rightarrow$ IT\} and \{IT $\rightarrow$ SP, CT\} also indicates high association with confidence range of between 89\% - 90\%. Similarly, \{SP $\rightarrow$ CT\}, \{IT $\rightarrow$ CT\}, \{IH $\rightarrow$ SP\} and \{IT, SP $\rightarrow$ CT\} are observed as the strongest rules with 100\% confidence in mobile applications as shown in Table \ref{table:assoc_rs}.
	As previously stated, the result of  the association is consistent with that of POPC algorithms represented in Figure \ref{fig:dendogram_rs}. In Figure \ref{fig:dendogram_rs}a (desktop application), we can observe a cluster \{SP, IT\} which coincides with the association rule \{IT $\rightarrow$ SP\} in Table \ref{table:assoc_rs}.
	
	\begin{table*}[h]%
		\caption{The association of the different role-stereotypes with each other across the desktop and mobile applications.\label{table:assoc_rs}}
		\begin{tabular*}{\textwidth}{@{\extracolsep\fill}lllll@{}}
			\toprule
			&\multicolumn{2}{@{}l}{\textbf{Desktop}} & \multicolumn{2}{@{}l}{\textbf{Mobile}} \\\cmidrule{2-3}\cmidrule{4-5}
			\textbf{No.} & \textbf{Rule}  & \textbf{Confidence}  & {\textbf{Rule}}  & \textbf{Confidence}   \\
			\midrule
			1. & IT $\rightarrow$ SP & 1.00 & SP $\rightarrow$ CT & 1.00 \\
			2. & ST  $\rightarrow$ SP & 1.00 & IT $\rightarrow$ CT & 1.00 \\
			3. & (IT, CT) $\rightarrow$ SP & 1.00 & IH $\rightarrow$ SP & 1.00 \\
			4. & CT  $\rightarrow$ SP & 0.90 & (IT, SP) $\rightarrow$ CT & 1.00 \\
			5. & IT $\rightarrow$ CT & 0.89 & IT $\rightarrow$ SP & 0.80 \\
			6. & (IT, SP) $\rightarrow$ CT & 0.89 & (IT, CT) $\Rightarrow$ SP & 0.80 \\
			7. & (SP, CT) $\rightarrow$ IT & 0.89 & (SP, CT) $\rightarrow$ IT & 0.80 \\
			8. & IT $\rightarrow$ (SP, CT) & 0.89 & IT $\rightarrow$ (SP, CT) & 0.80 \\
			\bottomrule
		\end{tabular*}
	\end{table*}
	
		\begin{tcolorbox}
		\faLightbulbO \hspace{2mm}\textbf{Summary of Results for RQ3:}
			\begin{enumerate}[leftmargin=*]
				\item \texttt{Service Provider} and \texttt{Information Holder} are prone to a wider range of design smells compared to other role-stereotypes such as \texttt{Coordinator} as shown in Table \ref{table:tb_highlighted}. 
				\item The \texttt{Interfacer} role-stereotype has sightly different types of design smells in desktop applications as compared to mobile applications. A hypothetical explanation is the availability of libraries and frameworks in the mobile mobile application development ecosystem.
				\item Figure \ref{fig:dendogram_rs} indicates a similarity in the following groups of role-stereotypes; \{\texttt{Coodinator}, \texttt{Structurer}\}, 
				\{\texttt{Controller}, \texttt{Interfacer}\} and 
				\{\texttt{Servicer Provider}, \texttt{Information Holder}\}. The Association rule closely agree with the clustering result.
			\end{enumerate}

	\end{tcolorbox}

	\section{DISCUSSIONS AND IMPLICATIONS}\label{sec:ds_imp}
	This section is divided into two subsections. In subsection \ref{discussion}, we discuss observations from our empirical study. The implications of this study are presented in subsection \ref{implications}.
	\subsection{Discussion of Results} \label{discussion}
	\subsubsection*{RQ1: How do design smells vary across mobile and desktop applications?}
	In addressing this research question, we explored the variations in design smells between mobile and desktop applications through two distinct methods. Initially, we compared the number of design smells per thousand lines of code (KLOC) for each selected project. As outlined in the results section, the study notes that the average number of design smells is higher in desktop applications compared to mobile applications. However, this disparity does not reach statistical significance. This finding aligns with earlier research by \textcite{mannan2016understanding}. It suggests that the prevalence of design smells in both desktop and mobile applications tends to be consistent, despite the distinct characteristics of these software ecosystems. Another reasonable explanation for this statistical result is the influence of developers' experience and coding styles, implying that less experienced developers are prone to introducing design smells in any project.
	
	Design smells co-occurrences have been identified through two unsupervised learning techniques capable of discovering frequent relationships in a dataset i.e. clustering and association rule mining, as illustrated in Figure \ref{fig:dendogram_ds} and Table \ref{table:assoc_ds_ds}, respectively. The results not only affirm some anticipated relationships but also unveil co-occurrences overlooked by prior research in the field. The clustering outcomes presented in Figure \ref{fig:dendogram_ds} offer practical validation of theories pertaining to shared characteristics and similarities among design smells. For instance, through unsupervised learning, we demonstrated that {\tt Speculative Generality} and {\tt SwissArmKnife} are closely related. Nevertheless, we also found some unexpected relationships or similarities in the clusters which require further research to comprehend and provide recommendations. Therefore, we encourage researchers to explore this direction in future studies. Additionally, our study establishes a solid foundation for software educators to illustrate various design principles to students. This allows learners to practically observe examples of both well-designed and poorly designed systems across a diverse range of software systems.
	
	\subsubsection*{RQ2: How do design smells relate with different types of role-stereotypes?}
	The findings in the results section outlined three major observations concerning this research question and this will inform our discussion as follows;- 
	
 \noindent As observed in Figure \ref{fig:pie_chart_percentages}, we believe that this result is associated with the characteristics and collaborative properties of role-stereotypes. For example, a {\tt Service Provider} is considered an object that performs specific work and offers services to other role-stereotypes on demand \cite{wirfs2006characterizing}. A typical {\tt Service Provider} class cache information and use it to improve performance or give clients more control over their operations \cite{wirfs2006characterizing}. The process of implementing all those operational logic can increase the chance of introducing design smells in source code. Specifically, {\tt LongMethod}, {\tt ComplexClass}, {\tt LongParameterList} and {\tt LazyClass} are identified as the main contributors to high percentage of design smells in the {\tt Service Provider} role-stereotype classes. A hypothetical explanation for the fact that {\tt LongMethod} is largely associated with {\tt Service Provider} is because services are based on long method definitions for API purposes.
 
    \noindent The {\tt Interfacer} role acts as a mediator to simplify communication with another system or subsystem. Specifically, It is responsible for handling and transforming requests and information between different parts of a system \cite{wirfs2006characterizing}. It is rare to have isolated code in today's enterprise software development. Software systems will often integrate with other systems such as payment APIs with an online shopping application, banking systems integrated with institutions' payment systems, or even subsystems interaction within your software. Based on the definition of a {\tt Interfacer} role, systems integration is one of the most difficult tasks in software development. As a result, software engineers often resort to performing several ``hacks'' in order to realise the desired solution while leveraging design smells (e.g. {\tt ComplexClass} and {\tt LongParameterList} as shown in Figure \ref{fig:bar_chart_percentage}) as a trade-off in the {\tt Interfacer} classes.
    \noindent We also noticed a relatively large distribution (15.1\%) of design smells in the {\tt Information Holder} role-stereotype classes. A basic example of Information Holder is the entity and value objects in a rich domain model. The relatively large amount of Information Holder across the selected projects can be related to the notion that an information holder may collaborate with service providers like data access classes or configuration classes to fetch more information on demand. 
    
    \noindent Overall, the selected projects have somewhat low distribution of design smells in the {\tt Structurer}, {\tt Controller} and {\tt Coordinator} role-stereotypes compared to the previously discussed class roles. This observation can be explained three-fold.
	First, there are few numbers of {\tt Structurer}, {\tt Controller} and {\tt Coordinator} role-stereotypes in the selected projects. This could imply that most software systems have few implementations of those class roles. However, more investigation should be conducted to support this claim. Secondly, it is likely that some class roles such as Service Provider are assuming too many responsibilities, in which case, a refactoring option should be explored. This could also be coupled with the nature of software systems, where, many functionalities are inclined towards services abstraction for API purpose,  transforming information and requests between software layer and data encapsulation. Software engineers and educators should investigate and design role-stereotype based refactoring options. Third, the characteristic of some role-stereotypes safeguards it from design smells. For example, a {\tt Coordinator} delegates work to other objects and it is not involved in a lot of decision making. Therefore, implementing a {\tt Coordinator} object does not involve complex logic and this can limit the possibility of design smells occurrence.
	
	\subsubsection*{RQ3: Does the type of application (desktop or mobile) influence the relation between design smells and role-stereotypes?}
	Although class responsibilities as categorized by \textcite{wirfs2006characterizing} are generic, there are some key notable differences that exist between mobile and desktop applications. For example, in desktop applications, the entry point is the {\tt main} method - this can be considered as `centalized flow of control'. In contrast, mobile (Android) applications do not depend on main methods but rather on event-handlers such as {\tt onCreate}, {\tt onResume}, etc. Hence, the overarching notion of behavioural interactions across components of mobile applications is `event-driven'.
	Furthermore,  desktop applications rely mostly on Java Swing library as their underlying Graphical User Interface (GUI) design library. However, this is not the case in mobile applications since there is a complete separation of the application logic from its presentation i.e. the  GUI  is mostly designed using eXtensible  Markup  Language (XML) \cite{mannan2016understanding}. Therefore, we believe that these differences could potentially influence the occurrence of design smells in various role-stereotypes.
	
The findings indicates that desktop applications are prone to design smells compared to mobile applications with the exception of the {\tt interfacer} role-stereotype as shown in Figure \ref{fig:bar_chart_percentage}. It is reasonable to believe that these variations are attributed to the underlying domain infrastructures - i.e. libraries and frameworks that are commonly used for developing applications. When developing mobile applications, certain libraries/frameworks  already exist and the developer does not have to rewrite the same logic from scratch hence reducing the possibility of introducing more design smells. Also, the mobile (Android) ecosystem has attracted a large community of developers and code reviewers who ensure high-quality code before release. However, the high amount of design smells in the {\tt Interfacer} role-stereotype of the mobile application is particularly interesting and merits further investigation. We think this can be explained in twofold; first, it indicates that a lot of functionality in mobile applications has to do with making API calls and requests between software layers. This is also supported by the fact that the Hardware Abstraction Layer (HAL) in Android OS provides standard interfaces that expose device hardware capabilities to the higher-level Java API framework\footnote{https://developer.android.com/guide/platform}. Secondly, the large community of code reviewers and developers can also be disadvantageous. \textcite{palomba2018beyond} observed that lack of communication or coordination between developers can cause community smells, making the code to be less maintainable and worse over time. Overall, it is sensible to think that the variety and number of parts that a system has might increase the occurrence of Interfacer-based design smells.
	
\subsection{Research Implications}\label{implications}
In this section, we present the implications of this study to software developers, tool creators and researchers.
	
\noindent \textbf{To Software Developers:} We discuss two fundamental implications for software developers in two main areas; (1) software design and development and (2) software quality assurance. The findings in the study provide important insights for effective software design and development. In essence, software system designers and developers can review class roles not only in term of their responsibility but also as an indicator of how vulnerability they are to certain group of design smells. This knowledge enables developers to pay extra attention to classes assigned with a certain responsibility. For example, Figure \ref{fig:pie_chart_percentages} and Figure \ref{fig:bubble_chart} indicate that classes which are classified as \texttt{Servicer Provider} are often more prone to a wide range of design smells such as \texttt{LongMethod}, \texttt{LongParameterList} and \texttt{ComplexClass}. Therefore, software developers should incorporate this information during the design-, Quality Assurance-phases and maintenance activities. Understanding which design smells commonly occur in a specific role-stereotype can help developers to look out for such design smells and quickly resolve them to improve the overall software quality and maintainability. Additionally, our results  can help develop or redefine already existing quality assurance guidelines and reduce the vulnerability of a system to design smells. Software quality assurance is an integral part of any software development and plays a significant role in ensuring that a software system conforms to a standard or predefined requirement. Good quality software eases maintenance and facilitates its evolution. 

As previously noted, our study revealed association of design smells with role-stereotypes. For example, we observed a high association of \texttt{LongMethod} with \texttt{Service Provider}. It seems likely that this association is inevitable or unavoidable, given the definition of \texttt{Service Provider}. Therefore, developers could ignore this design smell on condition that it is unavoidable and essential to their specific code. Moreover, not every smell is one a developer cares about or would fix. However, developers should be careful of \texttt{Speculative Generality} if the code is for a \texttt{Controller} role-stereotype.
	

\noindent \textbf{To Software Tool Builders:} Our finding confirms a recommendation by \textcite{nurwidyantoro2019automated}: it is likely to be beneficial to  tailor metrics for detecting design smells to specific role-stereotypes. The variation in type and magnitude noticed in the different role-stereotypes as shown in Figure \ref{fig:bubble_chart}, Figure \ref{fig:bar_chart_percentage} and Table \ref{table:tb_highlighted} is a clear indication to design smells detection tool creators that design smells metrics should not be applied uniformly across all classes irrespective of their roles. It would be better to optimize design smell metrics tailoring to a specific project when creating design smell detection tools. As observed by \textcite{kuzniarz2004empirical}, role-stereotypes are helpful for understanding UML class diagram. We believe that using the results obtained from this study, tool builders can begin to explore ways of automatically inferring design smells from UML class diagram and role-stereotypes information. It may even be possible to come up with good recommendations for refactoring for specific combinations of role-stereotypes and design smells.
	
\noindent \textbf{To Researchers:} The findings in the study indicate that researchers can study design smells from a role-stereotype perspective. This insight is significant for the creation of new knowledge and research direction within the scope of design smells and role-stereotype concepts. For example, researchers could explore the possibility of reverse engineering role-stereotype classification techniques to utilize design smells information observed in a given class. 
	Our results in Figure \ref{fig:bubble_chart}, Figure \ref{fig:dendogram_rs} and Table \ref{table:tb_highlighted} already sheds light on some of those preliminary features. This also implies that researchers studying role-stereotype can easily benefit from data provided by design smells research experts and vice versa, hence creating synergy between researchers studying design smells and role-stereotypes respectively. From a teaching and learning perspective, our findings provide dependable literature for software educator to demonstrator various theoretical concepts related to role-stereotypes and design smells. As such, learners can easily observe those concepts across a variety of software projects for the distribution of design smells across role-stereotypes as shown in Table \ref{table:tb_highlighted}.
	
	\section{THREATS TO VALIDITY}\label{sec:threats}
	In this section, we discuss the threats to validity of our work, such as possible faults in the tools employed, the generalizability, and repeatability of the presented results.
	
	\noindent \textbf{Construct validity: } measures the degree to which tools and metrics measure the properties they are supposed to measure \cite{sharma2021code}. It aims to ensure that observations and inferences made are appropriate based on the measurements taken during the study. In the context of understanding the relationship between role-stereotypes and design smells, the {\tt Ptidej} tool suite used in this study detects design smells based on some predefined set of metrics \cite{ogenrwot}. Relying on the outcome of this tool may pose a threat to validity since the metrics are static and assume that all design smells are equal in weight. However, we believe that the result of this tool is still valid for any static code analysis, which is the focus of our study. Another threat to construct validity is the use of  role-stereotypes classification replication package shared by the authors \cite{nurwidyantoro2019automated}. This tool was trained and evaluated on only one project, i.e., K9-Mail (773 Java classes). To mitigate this risk and improve the role-stereotypes classification accuracy, we re-trained the classification model with a larger dataset of over 5,000 Java classes.
	
	\noindent \textbf{Internal validity: }The study relied on two significant tools for building our fine-grained data, as previously mentioned. We used the Ptidej tool suite for the detection of design smells and the automated role-stereotypes classification replication package. We believe that the accuracy of our results also depends on the accuracy of those tools. In addition, to mitigate any multiplicity error that might result from the detection of both design smells and role-stereotypes, the study considered only classes with at least one type of design smell and participating in a role-stereotype. This is essential because for any given Java class, you might find a design smell but no/wrong role-stereotype, and vice versa, which can result in a large source of error and affect the overall accuracy.
	
	\noindent \textbf{External validity: }This concerns the generalizability and repeatability of the produced results. This study was carried out on software systems (desktop and mobile) written in the Java programming language only. However, various platforms within the mobile and desktop ecosystem exist, which also utilize object-oriented paradigms, and it would be important to further explore the relationship within these systems. Therefore, based on the shared characteristics and program structure of any OOP system, we believe that the methods used in this study can be replicated to benefit other object-oriented-based software. Another external threat to validity is the issue of using a relatively modest-sized dataset consisting of 30 projects (11,350 Java classes). Although we believe that the size of the dataset was modest to answer the research questions and draw conclusions, one could argue that using a larger-sized dataset would give more confidence to the results presented in this paper. Nonetheless, to encourage replication and future extensions of this work, we have developed an open-source replication package (scripts and dataset) available online.
	
	\noindent \textbf{Conclusion validity:} assesses the degree to which the conclusions we reached about the relationships in our data are reasonable. A low number of samples, which reduces the ability to reveal patterns in the data, is observed as one of the common threats to this type of validity \cite{wohlin2012experimentation}. Therefore, we aimed at achieving statistical reliability as much as possible, given our sample data. As discussed earlier, the Welch’s statistical test indicated that there is no significant difference between design smells in desktop and mobile applications in terms of the co-occurrence of design smells. The dataset used in this study consists of 11,350 Java class samples extracted from 30 OSS projects. Although we believe that the size of our dataset is considerable, using a larger-sized dataset would give more confidence to the results. Even so, our finding is still consistent with the work of previous authors \cite{mannan2016understanding}.
	
	\section{CONCLUSION}\label{sec:conclusion}
	
	In this paper, we have presented an exploratory study to understand the relation between design smells and role-stereotypes and how this relation  varies across desktop and mobile applications. We employed a number of statistical and unsupervised learning methods to report empirical evidence on the aforementioned relationships. Specifically, the study used two unsupervised learning methods i.e. clustering and association rule mining.
	
 Our findings shows that design smells tend to occur more often in \texttt{Service Provider}, \texttt{Interfacer} and \texttt{Information Holder} than in \texttt{Controller}, \texttt{Structurer} and \texttt{Coordinator} role-stereotypes. In addition, we found that design smells are more frequent in desktop than mobile applications especially in \texttt{Service Provider} and \texttt{Information Holder} role-stereotypes. Using clustering, this paper revealed that the following pairs \{\texttt{Coordinator}, \texttt{Structurer}\},\{\texttt{Controller}, \texttt{Interfacer}\} and \{\texttt{Service Provider}, \texttt{Information Holder}\} are often quite similar in terms of the type of design smells that often occur in them and within the ambit of desktop applications. This is in  comparison to \{\texttt{Interfacer}, \texttt{Service Provider}, \texttt{Information Holder}\} and \{\texttt{Coordinator}, \texttt{Controller}, \texttt{Structurer}\}  role-stereotypes in mobile applications. The results of this paper can guide software teams in their efforts to implement various code/design smell prevention and correction mechanisms, as well as improve and maintain conceptual integrity of classes during their design and maintenance. Moreover, much as we observed that certain design smells are associated to role-stereotypes (e.g. \texttt{LongMethod} and \texttt{Service Provider}), it could just as well be inevitable or unavoidable. In which case, we argue that a developer should ignore such design smells, since they could to some extent be unavoidable and essential to the code. Besides, not every design smell is one a developer cares about or would fix.
    
    \noindent \textbf{In future work}, we hope to extend this study to include an analysis of the introduction and evolution of design smells relative to role-stereotypes - i.e. are there any patterns over time? Besides, we believe that methodologies for system/architecture-design can benefit by using the lens of design smells across role-stereotypes. Therefore, it would be interesting to observe how different software architectures (microservice vs. monolithic) influence the occurrence of design smells for specific role-stereotypes. Using the findings of this study, It would be great to develop different tools to support developers in their day-to-day activities of software design and maintenance.



\section*{Data Availability}
The data and script that support the findings of this study are available in \href{https://doi.org/10.5281/zenodo.10775827}{10.5281/zenodo.10775827}

\printbibliography

\end{document}